\documentclass[12pt]{article} 

\usepackage{latexsym} 
\usepackage{amssymb}  
\usepackage{amsfonts} 
\usepackage{amsbsy}   
\usepackage{graphicx}       
\usepackage[dvips]{color}

\newcommand{\al}{\alpha}
\newcommand{\be}{\beta}
\newcommand{\ga}{\gamma}
\newcommand{\Ga}{\Gamma}
\newcommand{\de}{\delta}
\newcommand{\De}{\Delta}
\newcommand{\ep}{\varepsilon}

\newcommand{\ka}{\kappa}

\newcommand{\La}{\Lambda}

\newcommand{\si}{\sigma}
\newcommand{\Si}{\Sigma}
\renewcommand{\th}{\theta}   

\newcommand{\p}{\partial}
\newcommand{\<}{\langle} 
\renewcommand{\>}{\rangle} 

\newcommand{\txt}{\textstyle}

\newcommand{\dsp}{\displaystyle}

\newcommand{\ad}{\dagger}
\newcommand\eqn[1]{(\ref{#1})}      
\newcommand\Eqn[1]{Eq.~(\ref{#1})}  

\newcommand{\beq}{\begin{equation}}
\newcommand{\eeq}{\end{equation}}
\newcommand{\ba}{\begin{array}}
\newcommand{\bea}{\begin{eqnarray}}
\newcommand{\ea}{\end{array}}
\newcommand{\eea}{\end{eqnarray}}
\newcommand{\bc}{\begin{center}}
\newcommand{\ec}{\end{center}}
\newcommand{\ben}{\begin{enumerate}}
\newcommand{\een}{\end{enumerate}}

\newcommand{\dslash}{{\partial\kern-0.55em/}}
\newcommand{\Dslash}{{D\kern-0.65em/}}

\newcommand\comment[1]{ \hbox{[{\it Comment suppressed here.}\/]} }
\newcommand\hide[1]{}

\renewcommand{\O}{{\cal O}}

\newcommand{\skipover}[1]{}

\newcommand{\half} {{\txt \frac{1}{2}}}

\newcommand{\quarter}{{\txt \frac{1}{4}}}

\def\phm{\phantom{-}}

\makeatletter 

\def\section{
\setcounter{equation}{0}        
\@startsection {section}{1}{\z@}{-3.5ex plus -1ex minus 
 -.2ex}{2.3ex plus .2ex}{\large\bf}}
\renewcommand{\theequation}{\arabic{section}.\arabic{equation}}

\def\subsection{\@startsection{subsection}{2}{\z@}{-3.25ex plus -1ex minus 
 -.2ex}{1.5ex plus .2ex}{\normalsize\bf}}

\def\subsubsection{\@startsection{subsubsection}{3}{\z@}{-3.25ex plus
 -1ex minus -.2ex}{1.5ex plus .2ex}{\normalsize}}

\makeatother   

\newsavebox{\eqlabel}

\makeatletter  
\newlength{\numblen}
\newsavebox{\eqnumb}
\def\@eqnnum{\savebox{\eqnumb}{\rm (\theequation)}%
\settowidth{\numblen}{\usebox{\eqnumb}}%
\makebox[\numblen][l]{\usebox{\eqnumb}~~~\usebox{\eqlabel}}}
\makeatother   

\newenvironment{equationwithlabel}[1]{ %
  \begin{equation}\label{#1} }{\end{equation}} 
\newcommand{\beql}[1]{\begin{equationwithlabel}{#1}}
\newcommand{\eeql}{\end{equationwithlabel}}

\newcommand{\eV}{{\rm eV}} 
\newcommand{\keV}{{\rm keV}} 
\newcommand{\MeV}{{\rm MeV}} 
 
\newcommand{\vp}{{\mathbf p}}    
\newcommand{\vk}{{\mathbf k}}    

\newcommand{\pslash}{{p\kern-0.45em/}}
\newcommand{\kslash}{{k\kern-0.5em/}}

\newcommand{\thrbarA}{$\bar{\bf 3}_A$}
\newcommand{\sixS}{${\bf 6}_S$}
\newcommand{\thrS}{${\bf 3}_S$}
\newcommand{\oneA}{${\bf 1}_A$}

\newcommand{\psibar}{{\bar\psi}}

\newcommand{\cfms}{{\cal S}^{ij}_{\alpha\beta}}

\newcommand{\conds}[1]{\psi{#1}{\cal S}\psi}
\newcommand{\conda}[1]{\psi{#1}{\cal A}\psi}
\newcommand{\uu}{\mid\uparrow\uparrow\rangle}
\newcommand{\ud}{\mid\uparrow\downarrow\rangle}
\newcommand{\du}{\mid\downarrow\uparrow\rangle}
\newcommand{\dd}{\mid\downarrow\downarrow\rangle}

\begin{document}

\title{\bf Single color and single flavor\\ color superconductivity}

\author{
Mark G. Alford${}^{(a)}$, Jeffrey A. Bowers${}^{(b)}$, 
Jack M. Cheyne${}^{(a)}$, \\
Greig A. Cowan${}^{(a)}$ 
 \\[0.5ex]
\parbox{\textwidth}{
\parbox{0.9\textwidth}%
{
\normalsize
\begin{itemize}
\item[${}^{(a)}$] 
  Physics and Astronomy Department,
  Glasgow University,
  Glasgow, G12 8QQ, U.K.
\item[${}^{(b)}$] 
  Center for Theoretical Physics,
  Massachusetts Institute of Technology, Cambridge, MA 02139
\end{itemize}
}}
}

\newcommand{\preprintno}{
  \normalsize GUTPA/02/09/03, MIT-CTP-3315
}

\date{7 Oct 2002\\[1ex] \preprintno}

\begin{titlepage}
\maketitle
\def\thepage{}          

\begin{abstract}
We survey the non-locked color-flavor-spin
channels for quark-quark (color superconducting)
condensates in QCD, using an NJL model.
We also study isotropic quark-antiquark (mesonic) condensates.
We make
mean-field estimates of the strength and sign of the self-interaction
of each condensate, using four-fermion interaction vertices
based on known QCD interactions.
For the attractive quark pairing channels, we solve the
mean-field gap equations to obtain the size of the gap as a function of quark
density. We also calculate the dispersion relations for the
quasiquarks, in order to see how fully gapped the spectrum of
fermionic excitations will be.
We use our results to specify the likely pairing patterns 
in neutral quark matter,  and comment on possible phenomenological
consequences.
\end{abstract}

\end{titlepage}

\renewcommand{\thepage}{\arabic{page}}

\section{Introduction}
\label{sec:intro}

It is by now well known that the BCS mechanism that underlies
superconductivity in metals is likely to
operate even more strongly in dense quark matter
\cite{Barrois,BailinLove,IwaIwa,ARW2,RappEtc}
(for reviews, see Ref.~\cite{Reviews}).
Generally, pairing between quarks of different flavors has
received most attention. This is because the strongest
attractive interaction occurs in the color-antisymmetric spin-zero
channel, so by Fermi statistics the flavor wavefunction must
be antisymmetric, involving two different flavors.
This is the pattern of pairing in the heavily-studied
``2SC'' \cite{ARW2,RappEtc} and ``CFL'' \cite{ARW3} phases of quark matter.
However, there is strong reason to believe that these are not the
only pairing patterns that are relevant in nature, and in this
paper we survey and discuss some of the channels that are available
to quarks that cannot participate in 2SC or CFL pairing.

The essence of the BCS mechanism is that a Fermi surface
is unstable against pairing if there are any 
fermion-fermion channels in which the interaction is attractive \cite{BCS}.
For electrons in metals this condition is only met
where phonon-mediated attraction overwhelms 
Coulomb repulsion. But there are certainly channels
in which the dominant QCD
quark-quark interaction is attractive, as indicated by the fact
that at low density and temperature
quarks bind strongly together to form hadrons.
Mean-field calculations using QCD-inspired NJL models
\cite{ARW3,CarterDiakonov,BergesRajagopal,coupledgap},
lattice studies of NJL models \cite{3+1GrossNeveu},
and calculations using
gluon exchange with a hard dense loop resummed gluon propagator
\cite{Son,pert,rockefeller,SchaeferPatterns,ShovWij}
confirm this, and indicate that the main pairing patterns are
CFL (for sufficiently light strange quark or high density), or
2SC/2SC+s for heavier strange quarks.
However, these calculations do not take into account the requirement of
electric and color neutrality, which must be obeyed in
uniform dense matter in the real world, such as might be found in the
core of a compact star.

The effects of imposing neutrality were recently studied in
a model-independent  expansion in powers of $m_s/\mu$ \cite{AR-02}, and
in preliminary Nambu--Jona-Lasinio model calculations \cite{Steiner:2002gx}.
Those calculations indicate that the
2SC+s phase suffers a significant
penalty for remaining neutral, and may not occupy such a large
part of the phase diagram as previously thought. 

\begin{figure}[hbt]
\begin{center}
\includegraphics[width=0.9\textwidth]{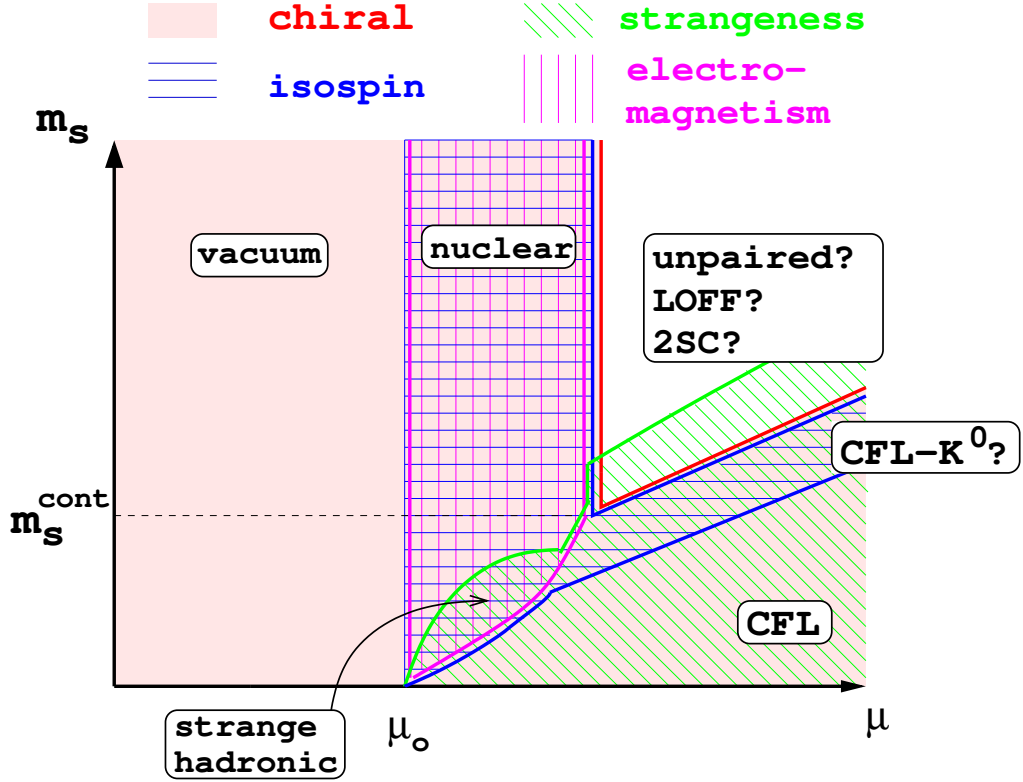}
\end{center}
\caption{
A conjectured phase diagram for neutral dense matter at zero temperature as a
function of quark chemical potential $\mu$ and strange quark mass
$m_s$ according to Ref.~\cite{AR-02}.  The up and down quark masses are
assumed zero.  The symmetries broken by the various phases are
indicated by the different shading styles.
The phase structure in the
chirally unbroken wedge in the upper right hand corner is still uncertain,
and is the topic of this paper.
}
\label{fig:phasediagram}
\end{figure}

A conjectured phase diagram
is sketched in Fig.~\ref{fig:phasediagram}, where
we show schematically the regions of
phase space in which various phases, each breaking a particular set of
symmetries, are expected to exist.
The diagram shows the $m_s$-$\mu$ plane
at temperature $T=0$, and  the $u$ and $d$ quark masses
are set to zero. 

At low strange quark mass ($m_s \ll m_s^{\rm cont}$), compression of
nuclear matter leads to the production of hyperons
(the ``strange hadronic phase''), and then a transition into 
an isospin-symmetric phase which by quark-hadron continuity \cite{SW-cont}
can be interpreted as a particular pairing pattern of baryons or as
quark matter in the chiral symmetry breaking color-flavor-locked phase.

At slightly higher
strange quark mass, compression of strange hadronic matter 
may lead into
an isospin broken phase that can be interpreted as hyperonic matter
with $\Si^0$ pairing \cite{Alford:2001mh} or as 
color-flavor-locked quark matter with K$^0$ condensation 
(``CFL-K$^0$'') \cite{BedaqueSchaefer}. However, the existence of this phase
at these densities is uncertain because of possibly 
large instanton effects \cite{Schaefer:2002gf}.

At high strange quark mass ($m_s>m_s^{\rm cont}$) the CFL phase
is pushed to higher densities, and there is an interval of densities
where we expect non-color-flavor-locked quark matter to exist.
This is the wedge in the upper right part 
of the $m_s$-$\mu$ plane (Fig.~\ref{fig:phasediagram}) where chiral symmetry
is restored. (In the figure it is assumed that close to
the transition to CFL there will be some strange quarks present, and they will
pair somehow, breaking strangeness.)
We are uncertain of the favored phase(s) in that region.
Since only the CFL phase pairs all the quarks that are present,
this region
will involve phases that leave single ``orphaned'' flavors or colors of
quark to find an attractive channel in which to pair. 
We can distinguish two general scenarios: \\
(1) There may be no BCS pairing of different flavors \cite{AR-02},
so all nine quark color/flavor
species are left to find alternative pairing channels.\\
(2) As the calculation of Ref.~\cite{Steiner:2002gx} suggests,
there may be some 2SC+s, i.e.~a phase that 
pairs the $u$ and $d$ quarks of two colors 
(red and green, say), leaving all the $s$ quarks
and the blue $u$ and $d$ quarks unpaired. \\
In both cases the BCS argument
implies that the supposedly unpaired quarks will in reality
seek {\em some} attractive channel in which to pair, 
and in this paper we will study some of the possibilities.
This question is of direct relevance to compact star physics,
where the smallest gap controls transport properties such
as the specific heat and neutrino emission rate.
The density of a compact star rises from nuclear density
near the surface to a much higher but as yet unknown value
in the core, so unless there is a direct transition from
nuclear to CFL quark matter, there will be regions where
some sort of single color or flavor pairing is likely.

One possibility for the orphaned quarks
is crystalline or ``LOFF'' pairing \cite{LOFF,OurLOFF,Bowers:2002xr}
in which species
whose Fermi surfaces are too far apart to support standard
translationally-invariant BCS pairing instead form pairs with net
momentum, utilizing only part of their Fermi surfaces, and
breaking translational invariance.
In this paper we will be concerned with another possibility:
single-color or single-flavor translationally invariant
(but not necessarily isotropic)
pairing \cite{IwaIwa,ARW2,TS1flav,BHO}. Such phases may have
non-zero angular momentum, spontaneously breaking rotational invariance.
In the following sections we survey the non-locked quark
pairing channels, paying particular attention to the
single flavor and/or single color channels. We identify the
attractive channels  using NJL models with four-fermion
interactions based on instantons, magnetic gluons, and
combined electric and magnetic gluons.
For the attractive channels we solve the gap equations,
deduce the quasiquark spectrum, and comment
on possible physical manifestations. Finally,
we perform a similar survey of mesonic channels, in which
exotic pairing has been posited \cite{zeroden}.

\section{Mean-field survey of quark pairing channels}
\label{sec:mean}

\subsection{Calculation}
To see which channels are attractive we perform a mean-field calculation
of the pairing energy for a wide range of condensation patterns.
We write the NJL Hamiltonian in the form
\beq
\label{hamiltonian}
\ba{rcl}
H &=& H_{\rm free} + H_{\rm interaction} \\[1ex]
H_{\rm free} &=& \psibar (\dslash - \mu\ga_0 + m) \psi \\[1ex]
 H_{\rm interaction} &=& {\psi^\dag}_{\al}^{ia} \psi^{\be b}_{j}
{\psi^\dag}_{\ga}^{kc} \psi^{\de d}_{l} 
\,\,{\cal H}^{\al}_{ia}{\,}_{\be b}^{j}{\,}^{\ga}_{kc}{\,}_{\de d}^{l}\ ,
\ea
\eeq
where color indices are $\al,\be,\ga,\de$, flavor indices are $i,j,k,l$,
spinor indices are $a,b,c,d$. The four-fermion interaction is
supposed to be a plausible model of QCD, so 
in the interaction kernel ${\cal H}$ we include three
terms, with the color-flavor-spinor structure of a two-flavor
instanton, electric gluon exchange, and magnetic gluon exchange,
\beq\label{kernel}
\ba{rcl}
{\cal H} &=& {\cal H}_{\rm elec} + {\cal H}_{\rm mag} + {\cal H}_{\rm inst}
  \\[1ex]
{\cal H}_{\rm elec} &=& \phm\frac{3}{8} G_E \,\,\de_i^j \de_k^l \,\,
  \de_{ab} \de_{cd} \,\,
  \frac{2}{3} (3 \de^\al_\de \de^\ga_\be - \de^\al_\be \de^\ga_\de) \\[1ex]
{\cal H}_{\rm mag} &=& \phm\frac{3}{8} G_M \,\,\de_i^j \de_k^l \,\,
  \sum_{n=1}^3 [\ga_0\ga_n]_{ab} [\ga_0\ga_n]_{cd} \,\,
  \frac{2}{3} (3 \de^\al_\de \de^\ga_\be - \de^\al_\be \de^\ga_\de) \\[1ex]
{\cal H}_{\rm inst} &=& -\frac{3}{4} G_I \,\,\ep_{ik} \ep^{jl} \,\,
  \quarter \Bigl( [\ga_0(1+\ga_5)]_{ab} [\ga_0(1+\ga_5)]_{cd} +
  [\ga_0(1-\ga_5)]_{ab} [\ga_0(1-\ga_5)]_{cd}
  \Bigr) \\
&& \phantom{-\frac{3}{4} G_I} 
   \frac{2}{3} (3 \de^\al_\be \de^\ga_\de - \de^\al_\de \de^\ga_\be)
\ea
\eeq

We consider condensates that factorize into separate color, flavor, and
Dirac tensors (i.e.~that do not show ``locking'')
and calculate their binding energy by contracting them with \eqn{kernel}.

There is no Fierzing ambiguity in this procedure.
For a given pairing pattern $X$, the condensate is
\beq
\label{qqcond}
\<\psi^{\be b}_{j}\psi^{\de d}_{l}\>^{\vphantom{y}}_{1PI} = \De(X)
  \mathfrak{C}_{(X)}^{\be\de} 
  \mathfrak{F}^{\vphantom{\be}}_{(X)jl} 
  \Ga_{(X)}^{bd\vphantom{\be}}~.
\eeq
We can then calculate the interaction (``binding'') energy of the various
condensates,
\beq
\label{qqstrength}
H = -\sum_X \De(X)^2 \Bigl(
   S^{(X)}_{\rm elec} G_E 
 + S^{(X)}_{\rm mag} G_M
 + S^{(X)}_{\rm inst} G_I \Bigr)
\eeq
The binding strengths $S^{(X)}_{\rm interaction}$ give the strength of
the self-interaction of the condensate $X$ due to the specified
part of the interaction Hamiltonian.

\subsection{Properties of the pairing channels}
In Table \ref{tab:channels} we list the the simple (translationally
invariant, factorizable)  channels available for quark pairing.
The meanings of the columns are as follows.
\ben
\item\underline{Color}:  
two quarks either make an antisymmetric color triplet
(which requires quarks of two different colors)  or
a symmetric sextet (which can occur with quarks of two different colors, 
and also if both quarks have the same color). 
For the \thrbarA\ we use 
$\mathfrak{C}^{\be\de}=\ep^{\be\de}$ in Eq.~\eqn{qqcond}. For the
\sixS\ we use a single-color representative
$\mathfrak{C}^{\be\de}=\de^{\be,1}\de^{\de,1}$ in Eq.~\eqn{qqcond}.
\item\underline{Flavor}: 
two quarks either make an antisymmetric flavor singlet
(which requires quarks of two different flavors) or
a symmetric triplet (which can occur with quarks of two different flavors, 
and also if both quarks have the same flavor).
For the \oneA\ we use $\mathfrak{F}_{jl}=
\si^2_{jl}$ and for the \thrS\ we use $\mathfrak{F}_{jl}=\si^1_{jl}$
in \eqn{qqcond}.
\item \underline{Spin,parity}: since the chemical potential explicitly breaks
the Lorentz group down to three-dimensional rotations and
translations, it makes sense to classify condensates by their total
angular momentum quantum number $j$ and parity.  
\item\underline{Dirac}:  This column gives the Dirac matrix structure
$\Ga^{bd}$ used in \eqn{qqcond}, 
so the condensate is  $\psi^T \Ga \psi$.
We also designate
each condensate as ``LL'' (even number of gamma matrices, so pairs
same-chirality quarks) or ``LR'' (odd number of gamma matrices, so
pairs opposite-chirality quarks). 
\item\underline{BCS-enhancement}: Condensates that correspond to pairs
of particles or holes near the Fermi surface have a BCS singularity
in their gap equation that guarantees a solution, no matter how
weak the coupling. To see which condensates have such a BCS enhancement,
we expanded the field operators in terms of creation and annihilation
operators (see Appendix \ref{app:angmom}). The order of the coefficient of the
$a(\vp)a(-\vp)$ and $b^\dag(\vp)b^\dag(-\vp)$ terms is given
in the table. $\O(1)$ means BCS-enhanced,
$0$ means not BCS-enhanced.
In the $C\ga_0\ga_5$ condensate the coefficient goes to zero as the quark
mass goes to zero (hence it is labelled ``$\O(m)$'' in the table)
meaning that the channel loses its BCS enhancement
in the chiral limit. This is discussed further in  Appendix \ref{app:angmom}.
\item\underline{Binding strength}: 
For each channel we show the binding strength for the instanton
interaction, the full (electric plus magnetic) gluon, 
which could reasonably be used at
medium density, and for the magnetic gluon alone,
which is known to dominate at ultra-high density \cite{BarroisPhD,Son}.
Channels with a positive binding strength
and BCS enhancement will always support pairing (the gap equation
always has a solution, however weak the couplings $G_I, G_E, G_M$).
Other things being equal, the pairing with the 
largest binding strength will have the lowest free energy, and
is the one that will actually occur.
\een

It may seem strange that there are entries in the table with
angular momentum $j=1$ and an antisymmetric Dirac structure
($C\ga_3\ga_5$), and with $j=0$ but a symmetric Dirac structure
($C \ga_0$). If all the angular momentum came from spin this
would be impossible. But even though there are
no explicit spatial derivatives in the diquark operators, there
can still be orbital angular momentum. In Appendix \ref{app:angmom}
the angular momentum content of the particle-particle component of the
condensates is analyzed into its
spin and orbital content. We see, for example, that
$C\ga_3\ga_5$ has an antisymmetric space wavefunction
($l=1$) and a symmetric spin wavefunction ($s=1$), combined
to give an antisymmetric $j=1$.

\subsection{Results}

\begin{table}[htb]
\setlength{\tabcolsep}{0.25em}
\newcommand{\attr}{\color{red}}
\newcommand{\rep}{\color{blue}}
\newcommand{\st}{\rule[-0.5ex]{0em}{2.8ex}}
\newlength{\instwid}
\newlength{\gluonwid}
\newlength{\magwid}
\settowidth{\instwid}{instanton} 
\settowidth{\gluonwid}{x+$S_{\rm mag}$x}
\settowidth{\magwid}{mag.~only}
\begin{tabular}{cccccccccccc}
\hline\hline
\multicolumn{6}{c}{Structure of condensate} 
 & \rule[-1ex]{0em}{3.5ex} &
\multicolumn{3}{c}{Binding strength} \\
\cline{1-6}\cline{8-10}
& & & & & & 
  & \rule[-2ex]{0em}{5ex} \makebox[\instwid]{instanton~} 
  & \multicolumn{2}{c}{gluon} \\[-1ex]
\cline{9-10}\\[-4.5ex]
color & flavor & $j$ & parity &
\multicolumn{2}{c}{Dirac} &  
\parbox{4em}{\bc BCS\\[-0.5ex] enhance-\\[-0.5ex] ment\ec}
 & \parbox{\instwid}{\bc ~\\ ~ \\[-0.2ex] $S_{\rm inst}$ \ec}
 & \parbox{\gluonwid}{
      \bc full\\ $S_{\rm elec}$\\[-0.2ex]+$S_{\rm mag}$ \ec}
 & \parbox{\magwid}{\bc mag.~only\\ ~ \\[-0.2ex] $S_{\rm mag}$ \ec } \\[-1ex]
\hline
\thrbarA & \oneA & $0_A$ & $+$& $C\ga_5$ & LL & $\O(1)$ 
  & \attr$+64$ & \attr$+64$ & \attr$+48$\st \\
\thrbarA & \oneA & $0_A$ & $-$& $C$ & LL & $\O(1)$  
  &  \rep$-64$ & \attr$+64$ & \attr$+48$ \\
\thrbarA & \oneA & $0_A$ & $+$& $C\ga_0 \ga_5$ & LR & $\O(m)$ 
  & 0 &  \rep$-32$ &  \rep$-48$  \\
\thrbarA & \oneA & $1_{A}$ &$-$& $C\ga_3 \ga_5$ & LR & $\O(1)$ 
  & 0 & \attr$+32$ & \attr$+16$  \\
\hline
\sixS & \oneA & $1_S$ &$-$& $C\si_{03} \ga_5$ & LL & $\O(1)$ 
  &  \rep$-16$ & 0 & \attr$+4$ \\
\sixS & \oneA & $1_S$ & $+$ & $C\si_{03}$ & LL & $\O(1)$ 
  &  \attr$+16$ & 0 & \attr$+4$ \\
\sixS & \oneA & $0_{S}$ &$-$& $C\ga_0$ & LR & 0  
  & 0 & \attr$+8$ & \attr$+12$ \\
\sixS & \oneA & $1_S$ & $+$ & $C\ga_3$ & LR & $\O(1)$ 
  & 0 &  \rep$-8$ &  \rep$-4$ \\
\hline
\thrbarA & \thrS & $1_S$ &$-$& $C\si_{03} \ga_5$ & LL & $\O(1)$ 
  & 0 & 0 &  \rep$-16$ \st \\
\thrbarA & \thrS & $1_S$ & $+$ & $C\si_{03}$ & LL & $\O(1)$ 
  & 0 & 0 &  \rep$-16$ \\
\thrbarA & \thrS & $0_{S}$ &$-$& $C\ga_0$ & LR & 0 
  & 0 &  \rep$-32$  &  \rep$-48$ \\
\thrbarA & \thrS & $1_S$ & $+$ & $C\ga_3$ & LR & $\O(1)$ 
  & 0 & \attr$+32$ & \attr$+16$ \\
\hline
\sixS & \thrS & $0_A$ & $+$ & $C\ga_5$ & LL & $\O(1)$ 
  & 0 &  \rep$-16$ &  \rep$-12$  \\
\sixS & \thrS & $0_A$ &$-$& $C$ & LL & $\O(1)$ 
  & 0 &  \rep$-16$ &  \rep$-12$  \\
\sixS & \thrS & $0_A$ & $+$ & $C\ga_0 \ga_5$ & LR & $\O(m)$  
  & 0 & \attr$+8$ & \attr$+12$ \\
\sixS & \thrS & $1_{A}$ &$-$& $C\ga_3 \ga_5$ & LR & $\O(1)$ 
  & 0 &  \rep$-8$  &  \rep$-4$  \\
\hline\hline
\end{tabular}
\caption{Binding strengths of diquark channels in NJL models
in the mean-field approximation.
The first 6 columns specify the channels,
and the last 3 columns give their attractiveness in NJL
models with various
types of four-fermion vertex: 2-flavor instanton, single gluon exchange,
single magnetic gluon exchange (expected to dominate at higher density).
See equations \eqn{qqcond} and \eqn{qqstrength} 
and subsequent explanation.
}
\label{tab:channels}
\end{table}

The results of the binding strength calculation are shown in Table
\ref{tab:channels}. The first block is antisymmetric in flavor and
color, and so describes pairing of two flavors and two colors. 
The second
block is for two flavors and one color, the third for one flavor and
two colors, and the final block is for one color and one flavor.

\begin{figure}[hbt]
\begin{center}
\includegraphics[width=0.9\textwidth]{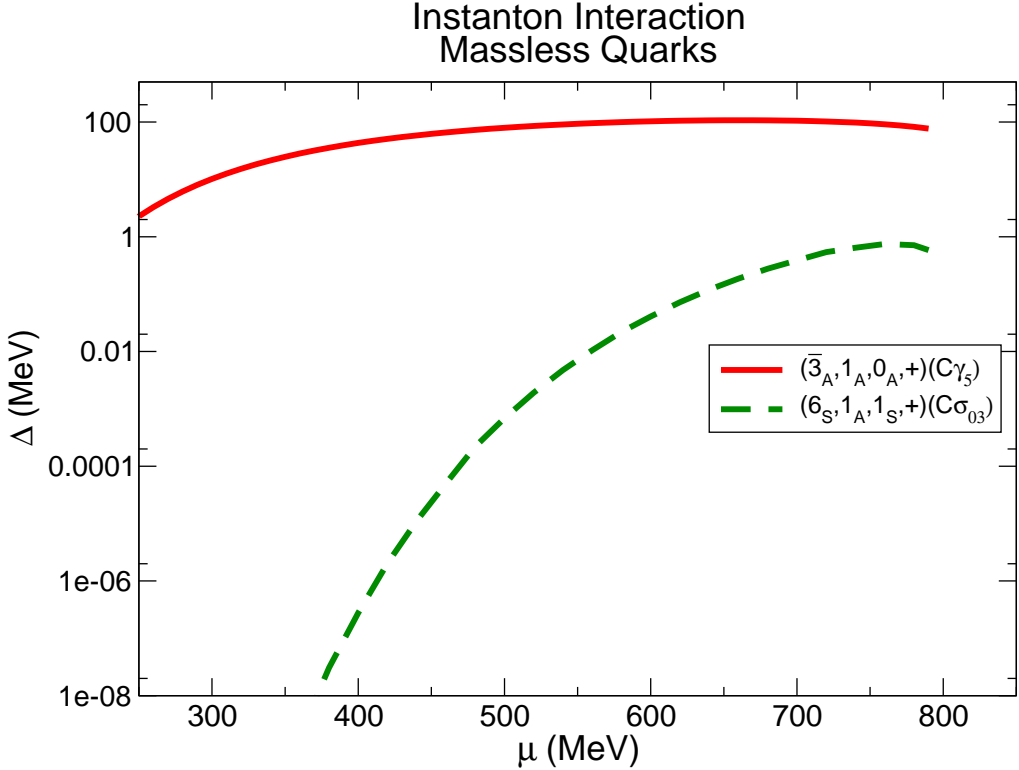}
\end{center}
\caption{Gap parameters in the attractive
channels for an NJL interaction based on the two-flavor instanton.
Since the instanton interaction requires two quark flavors, we
take the quarks to be massless, which is a good approximation for
the $u$ and $d$. The cutoff is $\La=800~\MeV$.}
\label{fig:Inst800}
\end{figure}

\begin{figure}[hbt]
\begin{center}
\includegraphics[width=0.9\textwidth]{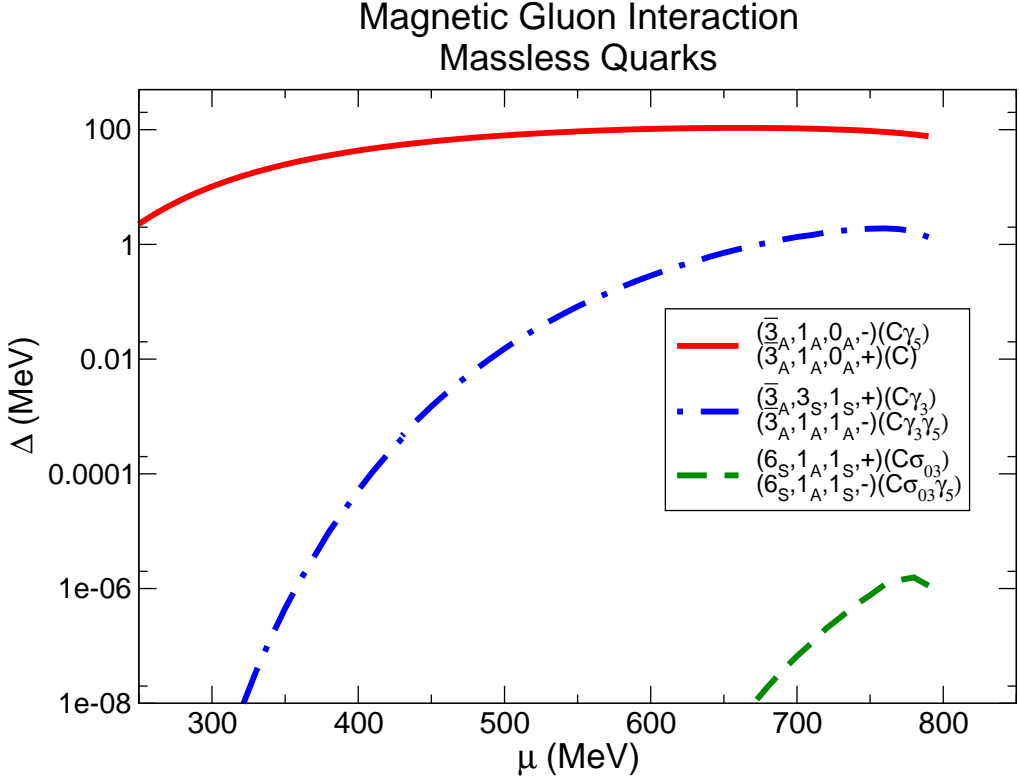}
\end{center}
\caption{Gap parameters in the attractive
channels for an NJL interaction based on 
magnetic-gluon exchange. We show the one-flavor and two-flavor channels,
for massless quarks.
The cutoff is $\La=800~\MeV$.
}
\label{fig:Mag800}
\end{figure}

\begin{figure}[hbt]
\begin{center}
\includegraphics[width=0.9\textwidth]{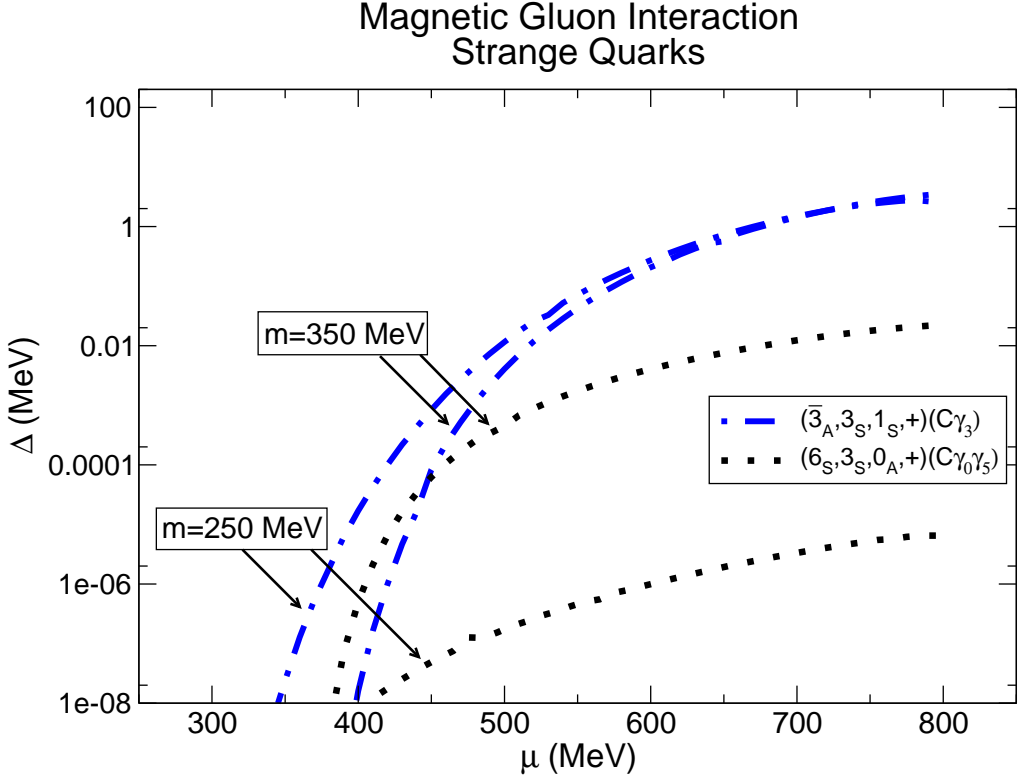}
\end{center}
\caption{Gap parameters in the attractive
channels for an NJL interaction based on magnetic-gluon exchange,
for quarks of mass 250 and 350~\MeV,
a reasonable range of values for the strange quark at medium density.
We show only the single-flavor channels. The cutoff is $\La=800~\MeV$.
}
\label{fig:Mag800strange}
\end{figure}

Certain features can be easily understood: the flavor-symmetric condensates
all have zero instanton binding energy, because the instanton
vertex is flavor-antisymmetric in the incoming quarks. 
The gluonic vertices give the same results for $C\ga_5$ as for $C$,
and for $C \si_{03}\ga_5$ as for $C \si_{03}$, because the gluonic
interaction is invariant under $U(1)_A$ transformations, 
under which the LL condensates
transform into each other ($C\ga_5 \rightleftharpoons C$ and
$C \si_{03}\ga_5 \rightleftharpoons C \si_{03}$) while the LR
condensates are invariant.
We see that there are many attractive channels:
\begin{list}{}{
 \setlength{\topsep}{-0.5\parskip}
 \setlength{\itemsep}{-0.5\parskip}
 }
\item[1)] Two colors and two flavors (\thrbarA,\oneA,$\ldots$).\\
The strongly attractive channel (\thrbarA,\oneA,0,$+$)($C\ga_5$) 
is the 2SC and CFL quark Cooper pairing pattern, and has been
extensively studied. The gap is large enough that even species 
with different masses, whose
Fermi momenta are quite far apart, can pair (hence the CFL phase
which pairs red and green $u$ and $d$, red and blue $u$ and $s$,
and green and blue $d$ and $s$ in this channel).
Its parity partner (\thrbarA,\oneA,0,$-$)($C$) is disfavored by instantons,
and is therefore unlikely to occur at phenomenologically interesting densities.
The additional channel
(\thrbarA,\oneA,1,$-$)($C\ga_3\ga_5$) is more weakly attractive
and also breaks rotational invariance,
and is therefore expected to be even less favored.
This is confirmed by gap equation calculations 
(Fig.~\ref{fig:Mag800}) which show that its gap is smaller 
by a factor of 10 to 100.
\item[2)] One color, two flavors (\sixS,\oneA,$\ldots$). \\
It is generally emphasized that the quark-quark interaction
is attractive in the color-antisymmetric \thrbarA\
channel. But, as we see in
table \ref{tab:channels}, the color-symmetric
(\sixS,\oneA,1,+)($C\si_{03}$) is attractive for instantons and
the magnetic gluon four-fermion interaction.
The instanton gives it a gap
of order 1~\MeV\ (Fig.~\ref{fig:Inst800}), while
the gluon interaction gives a small gap of order $1~\eV$
(Fig.~\ref{fig:Mag800}).
This channel was originally suggested for pairing of the blue
up and down quarks that are left out of 2SC \cite{ARW2},
and is discussed in more detail in Ref.~\cite{BHO}.
Its gap is small, so it could only pair quarks of similar mass,
i.e.~the light quarks, but
in a real-world uniform phase
such pairing will not occur either, because charge neutrality causes 
the up and down chemical potentials (and hence Fermi momenta)
to differ by tens of MeV, which is larger than the gap.
In a non-uniform mixture
of two locally charged phases \cite{Glendenning}, 
however, it is conceivable that
the up and down Fermi momenta could be similar enough to allow
pairing in this channel.
The parity partner (\sixS,\oneA,1,$-$)($C\si_{03}\ga_5$) is
disfavored by instantons. The channel 
(\sixS,\oneA,0,$-$)($C\ga_0$) is attractive, but 
has no particle-particle component, 
and presumably only occurs for sufficiently
strong coupling. Solving
the gap equations for reasonable coupling strength we find
no gap in this channel.
\item[3)] Two colors and one flavor (\thrbarA,\thrS,$\ldots$).\\
The only attractive channel is
(\thrbarA,\thrS,1,$+$)($C\ga_3$). 
This is a pairing option for
red and green strange quarks in 2SC+s.
We have solved the relevant gap equation
(Figs.~\ref{fig:Mag800},\ref{fig:Mag800strange}) 
and find gaps in the range $2-10~\MeV$.
If three colors are available then
a competing possibility is to lock the colors to the spin (CSL),
so the condensate is
a linear combination of $C \ga_i$ and $C \si_{0i}$
with a color structure that is correlated with the spatial direction,
e.g.~red and green quarks pair in the $z$ direction, red and blue
in the $y$ direction, green and blue in the $x$ direction.
This leaves an unbroken global $SO(3)$ of spatial rotations
combined with color rotations, so the gap is isotropic, which
helps to lower the free energy \cite{TS1flav,Schmitt:2002sc}.
Note also that the channels
(\thrbarA,\thrS,1,$+$)($C\si_{03}$) 
and (\thrbarA,\thrS,1,$-$)($C\si_{03}\ga_5$)
which are repulsive in the NJL model become
attractive at asymptotic density when 
the gluon propagator provides a form factor that
strongly emphasizes small-angle scattering  \cite{TS1flav}.
\item[4)] One color and one flavor (\sixS,\thrS,$\ldots$).\\
There is an attractive channel here, the
(\sixS,\thrS,0,$+$)($C\ga_0\ga_5$). 
It loses its particle-particle component as the quark mass
goes to zero, making it very weak for up and down quarks, but
quite strong for strange quarks (Fig.~\ref{fig:Mag800strange}).
It is suitable for the blue
strange quarks in 2SC+s when red and green strange quarks have
paired in the
(\thrbarA,\thrS,1,$+$)($C\ga_3$) channel.
\end{list}

Many of the attractive channels have repulsive partners
with the
same symmetries, so a condensate in the attractive
channel will automatically generate a small additional one in the
repulsive channel.
For example, the (\thrbarA,\oneA,0,$+$)($C\ga_5$) can
generate (\thrbarA,\oneA,0,$+$)($C\ga_0\ga_5$). This was discussed in
Ref.~\cite{ABR2+1}, where the induced ($C\ga_0\ga_5$) condensate
(there called ``$\ka$'') in 2+1 flavor CFL was calculated and found
to be small. In Ref.~\cite{Fugleberg:2002rk} it was observed that
if all three quarks are massive then this condensate may be important.
In the context of CFL the (\thrbarA,\oneA,0,$+$)($C\ga_5$) can
also generate (\sixS,\thrS,0,+)($C\ga_5$) \cite{ARW3,Pisarski:1999cn}, 
since they both break the full symmetry group down to the same subgroup.


\section{Gap calculations for the attractive diquark channels}
\label{sec:gap}

For the attractive channels we performed uncoupled gap equation calculations,
and obtained the dependence of the quark pairing on $\mu$. The
amount of pairing is given by the gap parameter $\De(\mu)$, which
occurs in the self energy (See Appendix \ref{app:gap}) as
\beq
\Delta^{\alpha\beta ab}_{ij}(p)=\Delta(\mu)\,\,{\cal C}^{\rm \alpha\beta}
  {\cal F}_{ij} \Gamma^{ab}~,
\eeq
with color matrix ${\cal C}$, flavor matrix ${\cal F}$, and Dirac
structure $\Gamma^{ab}$.
Note that $\De(\mu)$ is a gap {\em parameter}, not the gap. It sets
the scale of the gap in the quasiparticle excitation spectrum, but
as we will see in Section~\ref{sec:dispersion} the gap itself 
often depends on the direction of the momentum.

The 4-fermion interactions that we use are nonrenormalizable, so
our gap equation involves a 3-momentum cutoff $\La$, which
represents the decoupling of our interactions at higher momentum, due
to instanton form factors, effective gluon masses, etc.
The usual procedure for NJL model calculations is to calibrate
the coupling strength for each cutoff $\La$ by known low-density
physics such as the size of the chiral condensate. However, it is
well known that this leads to an approximately cutoff-independent
maximum gap (as a function of $\mu$) in the $\psi C\ga_5\psi$
channel, so we used that criterion directly as our calibration
condition, setting the maximum gap to $100~\MeV$.

The results of our calculations,
for cutoff $\La=800~\MeV$, are plotted in
Figs.~\ref{fig:Inst800},
\ref{fig:Mag800},\ref{fig:Mag800strange}. For other cutoffs the overall shape
of the curves is very similar. Because we use a sharp cutoff $\La$, the gap
falls to zero when $\mu$ reaches $\La$ (see, e.g., Eq.~\eqn{analyticgap}).
We show gap plots for the instanton interaction (Fig.~\ref{fig:Inst800})
and magnetic gluon interaction. The full electric + magnetic gluon gives
results that are similar to those for the magnetic gluon, but with
no gap in the (\sixS,\oneA,1,+)($C\si_{03}$) channel.

For the magnetic gluon, we show a gap plot for massless quarks
(Fig.~\ref{fig:Mag800}) which includes the two-flavor channels
(\thrbarA,\oneA,0,+)($C\ga_5$),  (\thrbarA,\oneA,1,$-$)($C\ga_3\ga_5$)
and (\sixS,\oneA,1,+)($C\si_{03}$)
which could sustain $u$-$d$ pairing, as well as the single-flavor
channel (\thrbarA,\thrS,1,+)($C\ga_3$) which could sustain $u$-$u$
or $d$-$d$ pairing.
We also show a gap plot for quarks with a mass appropriate to the 
strange quark (Fig.~\ref{fig:Mag800strange})
which includes the single-flavor channels that could sustain $s$-$s$
pairing with two colors (\thrbarA,\thrS,1,+)($C\ga_3$) 
or one color (\sixS,\thrS,0,+)($C\ga_0\ga_5$).

The relative sizes of the gaps in the different channels reflect the
pairing strengths given in Table \ref{tab:channels}. We see that the
Lorentz scalar (\thrbarA,\oneA,0,+)($C\ga_5$) (red/solid line) is dominant.
The $j=1$ channels have much smaller gap parameters.
The (\thrbarA,\thrS,1,+)($C \ga_3$) 
gap parameter (blue/dash-dot line in Figs.~\ref{fig:Mag800} 
and \ref{fig:Mag800strange})
rises to a few \MeV\ with the magnetic gluon interaction.
The (\sixS,\oneA,1,+)($C \si_{03}$) gap parameter
(green/dashed line) rises to about 1~\MeV\ with an instanton interaction,
but only 1~\eV\ with the magnetic gluon interaction. 
It should be remembered, however, that
the temperature of a compact star can be anything
from tens of MeV at the
time of the supernova to a few eV after millions of years, so
gaps anywhere in this range are of potential phenomenological interest.

The (\sixS,\thrS,0,+)($C\ga_0\ga_5$) channel (black/dotted line), 
which is the only attractive channel for
a single color and flavor of quark, is highly suppressed for
massless quarks at high density
but reaches about 10~\keV\ for strange
quarks ($m=350~\MeV$, Fig.~\ref{fig:Mag800strange}).
This is because its particle-particle component goes to
zero as $m\to 0$ (Eq.~\eqn{BCSenhancement} and Table~\ref{tab:channels}).

Up to this point we have not mentioned the $j=1, m_j=\pm 1$ channels
(e.g.~$\psi C\ga_\pm\psi \equiv \psi C(\ga_1\pm i\ga_2)\psi$).
We have only discussed the $j=1, m_j=0$
channels (e.g.~$\psi C\ga_3\psi$). That is because rotational invariance
of the interaction Hamiltonian that we are using
guarantees that changing $m_j$ from 0 to $\pm 1$ will not affect
the binding
energy and gap equation. This can be seen by considering the form of
the binding energy. From \Eqn{hamiltonian} it is
\beq
E_B \sim \<\psi\psi\>^{\dag ac} \<\psi\psi\>^{bd} {\cal H}_{abcd}~.
\eeq
Note that it is quadratic in the diquark condensate, with one of the
factors being complex conjugated.
So if we have some 3-vector condensate,
for example $\phi=\sum_i \phi_i \<\psi\ga_i \psi\>$, then its binding
energy is
\beq
E_B \propto |\phi_x|^2 + |\phi_y|^2 + |\phi_z|^2
\eeq
It is clear that the $m_j=0$ condensate $\phi_i=(0,0,1)$ has the same
binding energy as the $m_j=\pm 1$ condensate $\phi_i=(1/\sqrt{2})(1,\pm i,0)$.
We have explicitly solved the gap equations for the $m_j=\pm 1$
condensates, and find their solutions identical to the corresponding
$m_j=0$ condensates.
However, the quasiquark excitations in the two cases are quite different, 
and we proceed to study these in the next section.

\section{Quasiquark Dispersion relations}
\label{sec:dispersion}

The physical behavior of quark matter will be dominated by its lowest energy
excitations. As well as Goldstone bosons that arise from spontaneous
breaking of global symmetries, there will be fermionic excitations
of the quarks around the Fermi surface.
In the presence of a diquark condensate, the spectrum of quark
excitations is radically altered. Instead of arbitrarily low energy
degrees of freedom, associated with the promotion of a quark from a
state just below the Fermi surface to just above it, there is
a minimum excitation energy (gap), above which the excitation
spectrum is that of free
quasiquarks, which are linear combinations of a particle and a hole.

The dispersion relations of the quasiparticles can be calculated 
straightforwardly by including a condensate of the desired structure
in the inverse propagator $S^{-1}$, shown in Eq.~\eqn{invprop}.
Poles in the propagator correspond to zeros in $S^{-1}$, so the
dispersion relations are obtained by solving 
$\det S^{-1}(p_0,\vp,\mu,\De,m)=0$ for the energy $p_0$
as a function of the 3-momentum $\vp$ of the quasiparticle,
quark chemical potential $\mu$, gap parameter $\De$, and quark mass $m$.

\begin{figure}[hbt]
\begin{center}
\includegraphics[width=0.8\textwidth,angle=-90]{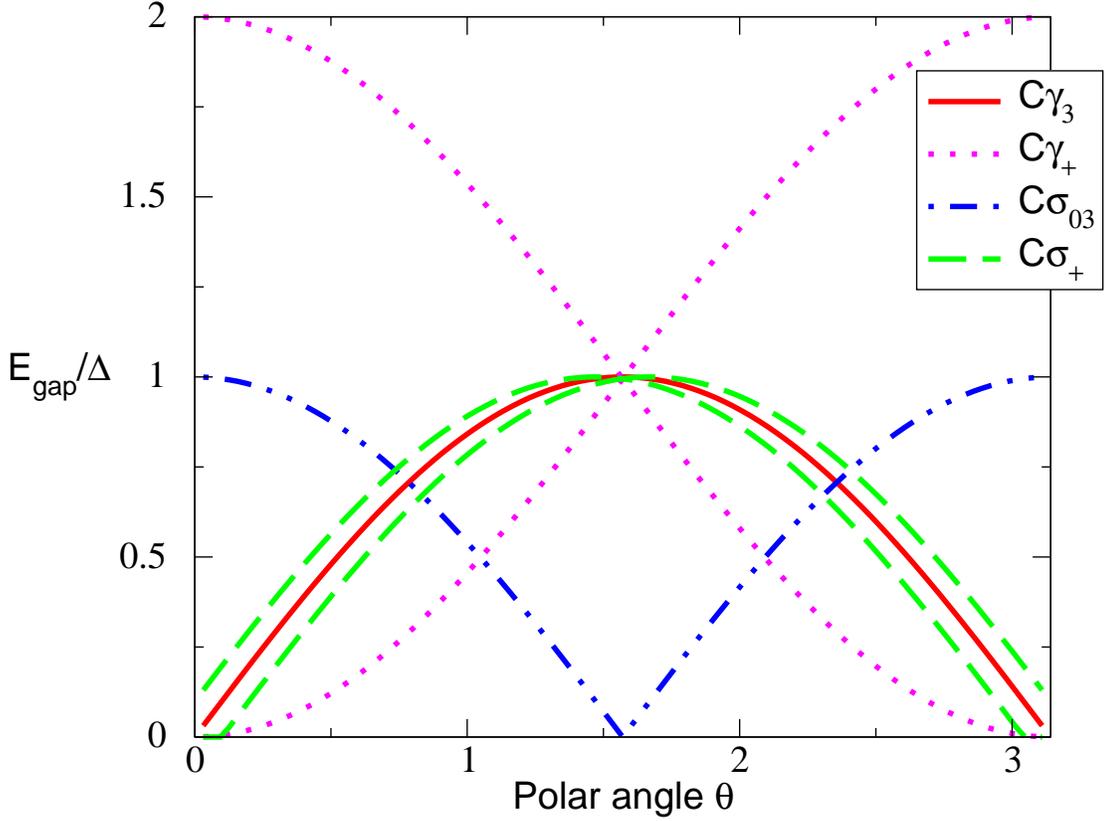}
\end{center}
\caption{Energy gap in units of the gap parameter
as a function of polar angle on the
Fermi surface for 
rotational symmetry breaking phases with
massless quarks, at $\mu=500~\MeV$, gap parameter $\De=50~\MeV$.
$\ga_+ \equiv \ga_1 + i\ga_2$, $\si_+ \equiv \si_{01} + i\si_{02}$.
}
\label{fig:DRmassless}
\end{figure}

\begin{figure}[hbt]
\begin{center}
\includegraphics[width=0.8\textwidth,angle=-90]{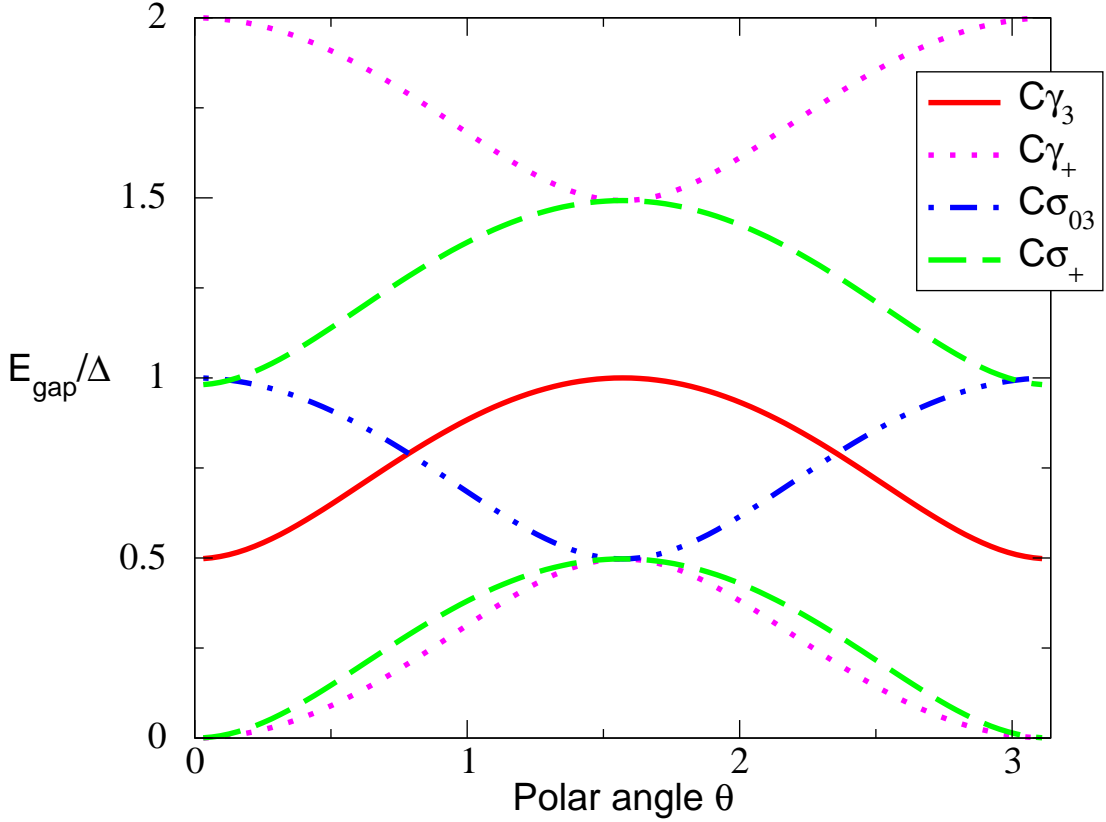}
\end{center}
\caption{Energy gap in units of the gap parameter
as a function of polar angle on the
Fermi surface for
rotational symmetry breaking phases at 
$\mu=500~\MeV$, with gap parameter $\De=50~\MeV$.
The quarks have mass $m=250~\MeV$.
$\ga_+ \equiv \ga_1 + i\ga_2$, $\si_+ \equiv \si_{01} + i\si_{02}$.
}
\label{fig:DRmassive}
\end{figure}

The gap is by definition the energy required to excite
the lowest energy quasiquark mode. Isotropic condensates have a uniform
gap, but one of the most interesting features of $j> 0$ condensates
is that they are not in general fully gapped: the gap goes to zero
for particular values of momentum $\vp$, which correspond to
particular places on the Fermi surface. This means that
transport properties such as viscosities and emissivities, 
which are suppressed by factors of $\exp(-\De/T)$ in phases
with isotropic quark pairing, may not be so 
strongly suppressed by a $j>0$ condensate.
In Figs.~\ref{fig:DRmassless} and \ref{fig:DRmassive} we show the 
variation of the gap over the Fermi surface by plotting
the energy of the lowest excitation as a function of angle, 
\beq
E_{\rm gap}(\th) = \min_{p,i} |E_i(p,\th)|
\eeq
where $E_i(p,\th)$
is the energy of the $i^{\rm th}$ quasiquark excitation
with momentum $(p\sin\th\cos\phi, p\sin\th\sin\phi, p\cos\th)$.
For the plots we take $\mu=500~\MeV$ and $\De=50~\MeV$,
with quark mass $m=0$ (Fig.~\ref{fig:DRmassless}) or
$m=250~\MeV$ (Fig.~\ref{fig:DRmassive}).

\newcommand{\mueff}{\mu_{\rm eff}}
\newcommand{\Deltaeff}{\De_{\rm eff}}

\begin{list}{\arabic{enumi})}{
  \usecounter{enumi}
  \setlength{\itemsep}{-0.5\parsep} 
  \setlength{\labelwidth}{0mm}
  \setlength{\labelsep}{0mm}
  \setlength{\leftmargin}{0 mm} 
 }
\item ~~{$C\ga_3$} condensate: $j=1,m_j=0$.\\
There is one quasiquark excitation with
energy less than the gap parameter $\De$.
\beq
\ba{rcl}
E(p)^2 &=& (\sqrt{p^2 + m^2 \mu^2/\mueff^2}\pm \mueff)^2 + \Deltaeff^2\\[1ex]
\mueff(\th)^2 &=& \mu^2 + \Delta^2 \cos^2(\theta) \\[1ex]
\Deltaeff(\th)^2 &=& \Delta^2 \Bigl( 
  \sin^2(\theta) + m^2/\mueff^2 \cos^2(\theta) \Bigr)
\ea
\eeq
From the expression for $\Deltaeff(\th)$ we see that for massless quarks
the gap goes to zero for momenta parallel to the $z$-axis, i.e.~at the
poles on the Fermi surface (red/solid curve in Fig.~\ref{fig:DRmassless}). 
Massive quarks retain a small gap of order
$ m\De/\mu$ at the poles (red/solid curve in Fig.~\ref{fig:DRmassive}).

\item ~~$C(\ga_1\pm i\ga_2)$ condensate: $j=1,m_j=\pm 1$.\\
There are two quasiquark excitations with
energy less than $2\De$,
\beq
\ba{rl}
E(p)^2 = 2\Delta^2 + m^2+ \mu^2 + p^2 \pm &
  \Bigl( 4\De^4 + 4\mu^2(p^2+m^2) +  2\De^2 p^2(1 - \cos(2\th)) \\ 
  & \pm 4\De^2\mu \sqrt{4 m^2 + 2 p^2(1+\cos(2\th))} \Bigr)^\half
\ea
\eeq
For this condensate the effective gap again goes to zero at the poles,
but in this case it remains zero even in the presence of a quark mass
 (magenta/dotted curve in
Figs.~\ref{fig:DRmassless},\ref{fig:DRmassive}).

\item ~~$C\si_{03}$ condensate:  $j=1,m_j=0$.\\
There is one quasiquark excitation with energy less than $\De$,
its dispersion relation is \cite{BHO}
\beq
\ba{rcl}
E(p)^2 &=& (\sqrt{p^2 + m^2 \mu^2/\mueff^2}\pm \mueff)^2 + \Deltaeff^2\\[1ex]
\mueff(\th)^2 &=& \mu^2 + \Delta^2 \sin^2(\theta) \\[1ex]
\Deltaeff(\th)^2 &=& \Delta^2 \Bigl( 
  \cos^2(\theta) + m^2/\mueff^2 \sin^2(\theta) \Bigr)
\ea
\eeq
This is related to the dispersion relation for the 
$C\ga_3$ condensate by $\th \to \pi/2-\th$: for massless quarks the
quasiquarks are gapless around the equator of the Fermi sphere
(blue/dash-double-dot curve in Fig.~\ref{fig:DRmassless})
and in the presence of a quark mass they gain a small gap of order
$\De m/\mu$ (blue/dash-double-dot curve in Fig.~\ref{fig:DRmassive}). 
The equator is a larger proportion of the Fermi surface than
the poles, so in this case we might expect a greater effect
on transport properties.

\item ~~$C(\si_{01}\pm i\si_{02})$ condensate: $j=1,m_j=\pm 1$.\\
There are two quasiquark excitations with 
energy less than $\De$. They have
rather complicated dispersion relations. 
Going to the massless
case, and assuming $E,(p-\mu) \ll \mu$, which will be true for the low-energy
quasiquark degrees of freedom that we are interested in, we find
\beq
\ba{rcl}
E(p) &=& (\De^2 \mu + \De^2 p \cos(\th) \pm \eta/\sqrt{2})/(2\mu^2) \\
\eta^2 &=& ( 8\mu^4(\mu - p)^2 + 8\De^2\mu^2(\mu^2 - \cos(\th)^2 p^2) +
  2\De^4(\mu + \cos(\th) p)^2 )
\ea
\eeq
In this case there is a region near the poles, $\th \lesssim \De/\mu$,
where the gap is zero (green/dashed curve in Fig.~\ref{fig:DRmassless}). 
This is because at those angles $E(p)$ has
zeros at  two values of $p$ close to $\mu$. When $\th\approx \De/\mu$
those two zeros merge and disappear from the real $p$ axis.
The presence of a quark mass $m>\De$ wipes out this effect, but there
is still no gap at the poles on the Fermi surface (green/dashed curve
in Fig.~\ref{fig:DRmassive}).
\end{list}

In comparing Figs.~\ref{fig:DRmassless} and \ref{fig:DRmassive}
it is interesting to note that introducing a mass for the quark
opens up a gap whenever the  gap lines intersect each other
at a non-zero angle (after one includes the mirror-image negative-energy
gap curves for the quasiholes). This occurs at zero energy at the poles
for $C\ga_3$ and at the equator for $C\si_{03}$. It occurs at
non-zero energy for $C(\ga_1\pm i\ga_2)$. The case of 
$C(\si_{01}\pm i\si_{02})$ is similar, but it is not obvious from the
gap plot for reasons described above.

We see that the $j\neq 0$ phases show a rich variety of
quasiquark dispersion relations. 
For massless quarks they are all gapless in special regions
of the Fermi surface, and for massive quarks the $m_j=\pm 1$
condensates remain gapless for momenta parallel to the spin.
It follows that for these phases the quasiquark excitations will
play an important role in transport properties, even
when the temperature is less than the gap parameter.

Moreover, different condensates ($m_j=\pm 1$ vs.~$m_j=0$) that 
because of rotational invariance of the Hamiltonian have
exactly the same binding energy and gap equation, nevertheless
have completely different energy gaps over the Fermi surface.
They will therefore behave quite differently when exposed
to nonisotropic external influences, such as magnetic fields
or neutrino fluxes, and also in their coupling to external
sources of torque, e.g.~via electron-quasiquark scattering.
All these influences are present in compact stars, and it will
be interesting and complicated to sort out which is favored
under naturally occurring conditions. And it should not
be forgotten that these conditions vary with the age
of the star. 

\section{Mesonic condensates}
\label{sec:mesonic}

Berges and Wetterich \cite{zeroden} have invoked ideas of
complementarity to suggest that the confining QCD vacuum could
be understood in terms of non-color-singlet chiral
condensates.  We can study the attractiveness of such
channels by mean-field methods similar to those of section
\ref{sec:mean}.

A mesonic condensate $X$ with strength $A_{(X)}$ is
\beq
\<\psibar_{\al a}^{i}\psi^{\be b}_{j}\> = A_{(X)}\,
  {\mathfrak{CF}_{(X)}}^{\be i}_{\al j}\, {\Ga_{(X)}}_a^b
\label{qbarcond}
\eeq
where $\mathfrak{CF}$ is the color-flavor structure of the
mesonic condensate.
We can
calculate their binding energy by contracting them with \eqn{kernel},
and the energy is given by \eqn{qqstrength}, where as before the
binding strengths $S^{(X)}_{\rm interaction}$ give the strength of
the self-interaction of the condensate $X$ due to the specified
part of the interaction Hamiltonian. Unlike the diquark condensates,
there are two possible contractions, the Hartree contribution
and the Fock contribution, since there are two possible
$\psi^\ad$ to contract with each $\psi$. Again there is
no Fierz ambiguity, and we include both Hartree and Fock
contributions.

For consistency we have used the same three interactions, 
namely instanton,
full (electric plus magnetic) gluon, and magnetic gluon,
as in our treatment of quark pairing. However,
it should be noted that the main relevance of quark-antiquark
pairing is at low density, where there is no reason to expect the
magnetic gluon to predominate, so in the case of Table~\ref{tab:mesonic}
the final column does not have any special physical relevance.

The color-flavor structures that we study are
\beq
\ba{rcl}
 {\mathfrak{CF}_{({\bf 8},{\bf 8})}}^{\al j}_{\be i} &=& 
   \de^\be_j \de_\al^i - \half \de^\be_\al \de_j^i \\[0.5ex]
 {\mathfrak{CF}_{({\bf 1},{\bf 1})}}^{\al j}_{\be i} &=& 
   \de_j^i \de^\be_\al \\[0.5ex]
 {\mathfrak{CF}_{({\bf 8},{\bf 1})}}^{\al j}_{\be i} &=&
  \de_j^i ( \de^\be_\al \de^\be_1 + \de^\be_\al \de^\be_2
 -2 \de^\be_\al \de^\be_3)
\ea
\eeq
The ({\bf 1},{\bf 1}) is a singlet in color and flavor.
The ({\bf 8},{\bf 8}) is a color-flavor locked adjoint 
chiral condensate, of the kind posited in Ref.~\cite{zeroden}.
We have studied it in the two-flavor case, where the
two flavors lock to two of the three colors (red and green, say).
The ({\bf 8},{\bf 1}) is a flavor-singlet condensate with
color structure $(1,1,-2)$, so it spontaneously selects
a color direction that we have fixed as blue.

The Dirac structures are two chiral condensates $1$ and $\ga_5$
(scalar and pseudoscalar respectively), and two that contain a $\ga_0$
and therefore correspond to a number density of quarks.
 $N_R+N_L$ for
$\ga_0$, i.e.~a renormalization of the chemical potential, and $N_R-N_L$
for $\ga_0\ga_5$.

\begin{table}[htb]
\newcommand{\attr}{\color{red}}
\newcommand{\rep}{\color{blue}}
\settowidth{\magwid}{magnetic}
\newlength{\repwid}
\settowidth{\repwid}{Structure of}
\begin{center}
\begin{tabular}{cc@{~~~~~}ccc}
\hline\hline
\multicolumn{2}{c}{\rule[-1ex]{0em}{4ex}
   \parbox{\repwid}{Structure of \\[-0.5ex] condensate}} &
\multicolumn{3}{c}{Attractiveness} \\
\cline{3-5}
color-flavor & Dirac
  & instanton 
  & \parbox{6em}{\bc mag + elec\\[-0.5ex] gluon \ec} 
  & \rule[-2ex]{0em}{4ex}
  \parbox{\magwid}{\bc magnetic\\[-0.5ex] gluon \ec} \\
\hline
({\bf 1},{\bf 1}) & $1$     & \attr$+192$  & \attr$+192$ & \attr$+144$ \\
({\bf 1},{\bf 1}) & $\ga_5$ & \rep$-192$ & \attr$+192$ & \attr$+144$ \\
({\bf 1},{\bf 1}) & $\ga_0$       & 0 & \rep$-96$ & \rep$-144$ \\
({\bf 1},{\bf 1}) & $i\ga_0\ga_5$ & 0 & \rep$-96$ & \rep$-144$ \\
\hline
({\bf 8},{\bf 8}) & $1$           & \rep$-6$ & \rep$-12$ & \rep$-9$ \\
({\bf 8},{\bf 8}) & $\ga_5$       & \attr$+6$ & \rep$-12$ & \rep$-9$ \\
({\bf 8},{\bf 8}) & $\ga_0$       & 0 & \attr$+6$ & \attr$+9$ \\
({\bf 8},{\bf 8}) & $i\ga_0\ga_5$ & 0 & \attr$+6$ & \attr$+9$ \\
\hline
({\bf 8},{\bf 1}) & $1$           & \attr$+24$ & \rep$-48$ & \rep$-36$ \\
({\bf 8},{\bf 1}) & $\ga_5$       & \rep$-24$ & \rep$-48$ & \rep$-36$ \\
({\bf 8},{\bf 1}) & $\ga_0$       & 0 & \rep$-300$  & \attr$+36$ \\
({\bf 8},{\bf 1}) & $i\ga_0\ga_5$ & 0 & \rep$-12$  & \attr$+36$ \\
\hline\hline
\end{tabular}
\end{center}
\caption{Quark-antiquark pairing strength in various 
rotationally symmetric channels.
}
\label{tab:mesonic}
\end{table}

The results are given in table \ref{tab:mesonic}.
We see that the standard chiral condensate $({\bf 1},{\bf 1},1)$
is overwhelmingly favored over all the others. 
Its pseudoscalar partner 
({\bf 1},{\bf 1})$(\ga_5)$ is disfavored by instantons.

Among the color-flavor-locked adjoint condensates, 
the interaction energies are much weaker. The Lorentz scalar
({\bf 8},{\bf 8})$(1)$
studied in Ref.~\cite{zeroden} is disfavored by all the interactions,
but its pseudoscalar
partner ({\bf 8},{\bf 8})$(\ga_5)$ is less disfavored,
thanks to a positive contribution from instantons.
Interestingly, the only favored adjoint condensates are the
ones that correspond to spontaneous generation of particle number,
the ({\bf 8},{\bf 8})$(\ga_0)$ and its pseudoscalar partner.

The color-adjoint flavor-singlet condensates show larger interaction
energies than the color-flavor-locked adjoint condensates, but mostly they
are repulsive. The very large repulsion for the ({\bf 8},{\bf 1})$(\ga_0)$
in the electric gluon channel comes from the Hartree term, and
corresponds to the enormous Coulomb repulsion of a 
spontaneously generated uniform density of blue quarks. This 
color imbalance does not arise in other condensates:
in the color-flavor locked ({\bf 8},{\bf 8})$(\ga_0)$ each flavor
favors a different color, and in the ({\bf 8},{\bf 8})$(i\ga_0\ga_5)$
the left-handed and right-handed quarks have opposite color.
If instantons predominate then the ({\bf 8},{\bf 1})(1) condensate
would be the strongest competitor to the standard chiral condensate.


\section{Summary}
We have surveyed the factorizable (non-locked) color-flavor-spin
quark pairing condensates that occur in an NJL model of QCD that
uses point-like four-fermion interactions (Table \ref{tab:channels}). 
We have solved the gap equation
for the attractive channels, and obtained the relative sizes of their
gap parameters 
(Figs.~\ref{fig:Inst800},\ref{fig:Mag800},\ref{fig:Mag800strange}).

We conclude that there are three types of neutral
quark matter that may be found
in the mysterious non-CFL wedge in the 
high strange quark mass and high density region
of the $m_s$-$\mu$ plane
(Fig.~\ref{fig:phasediagram}): two-flavor pairing of
$u$ and $d$ quarks with additional 
single-flavor pairing of the remaining species (``2SC+1SC''),
single-flavor pairing only (``1SC''), and
three-flavor crystalline (``LOFF'') pairing.

These phases arise out of the competition between three factors:
the strange quark mass, which increases the energy cost
of $s$ quarks, depressing their Fermi momentum;
the requirement of electrical neutrality, which increases
the number of $u$ quarks to compensate for the smaller number
of $s$ quarks; the energy gain that quarks obtain by pairing.

\begin{figure}[hbt]
\begin{center}
\includegraphics[width=0.9\textwidth]{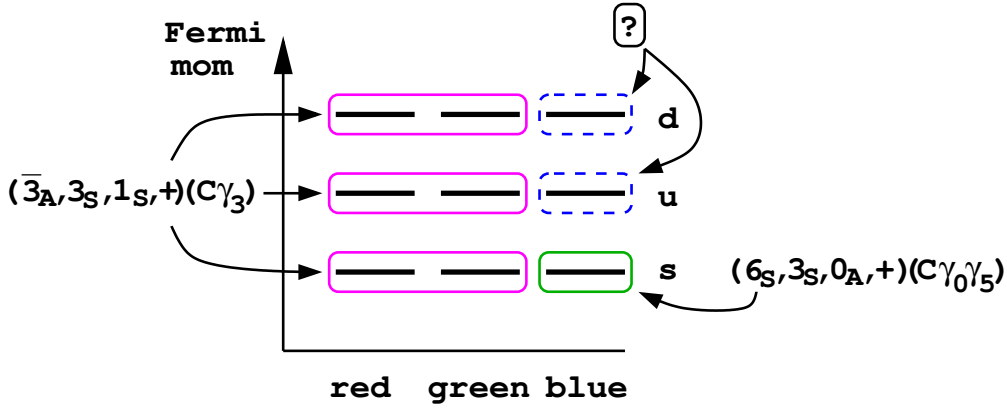}
\end{center}
\caption{Pictorial representation of single flavor pairing (``1SC'')
in neutral quark matter. 
The requirement of electric neutrality and a nonzero strange quark mass
forces the Fermi momenta of the three flavors apart. The
red and green colors of each flavor pair in a color-antisymmetric channel.
The blue $s$ quark self-pairs in a color and flavor symmetric channel.
We have not found an attractive channel in which the blue $u$ or
blue $d$ could pair. Competing pairing patterns for the same arrangement
of Fermi momenta are color-spin locking and the CFL-LOFF phase (see text).
}
\label{fig:unpaired_levels}
\end{figure}

\begin{figure}[hbt]
\begin{center}
\includegraphics[width=0.9\textwidth]{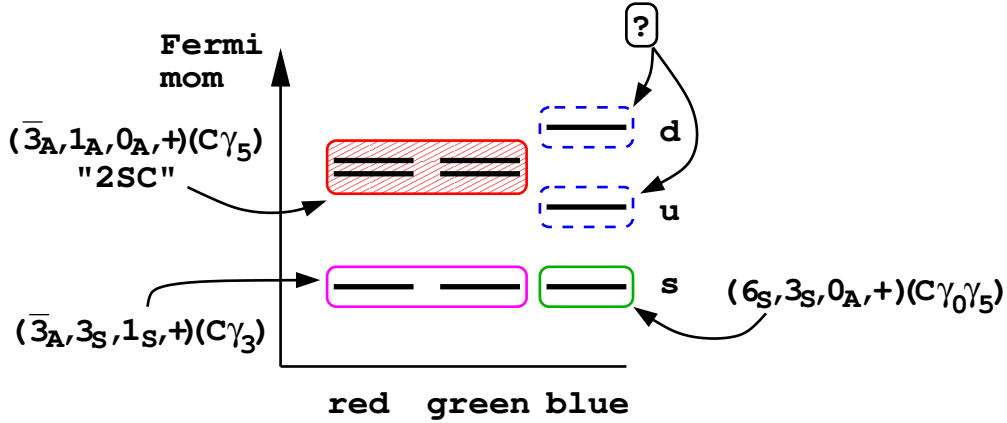}
\end{center}
\caption{Pictorial representation of 2SC+s with
additional single-flavor pairing
in neutral quark matter. This will only occur if the condensation energy
of the 2SC pairing is strong enough to offset the cost of dragging
the red and green $u$ and $d$ Fermi momenta away from the values
dictated by electric neutrality and the strange quark mass, 
to a common value.
}
\label{fig:2sc_levels}
\end{figure}

\ben
\item Single-flavor pairing (1SC).\\
If the strange quark mass is large enough then
charge neutrality requirements will
force all three flavors to have Fermi momenta
too far apart to allow any BCS pairing of different flavors.
In this case,
each flavor will have to self-pair (Fig.~\ref{fig:unpaired_levels})%
\footnote{
For a strange quark mass that is large
(probably unphysically so), there will
be a region in the quark matter part of the $m_s$-$\mu$ plane where 
$\mu<m_s$, so there are no $s$ quarks and the $u$ and $d$ 
Fermi momenta differ by some appreciable fraction of $\mu$. 
This region is described by
Fig~\ref{fig:unpaired_levels} or Fig~\ref{fig:2sc_levels}
but with the strange quark Fermi momentum set
to zero, so no pairing of strange quarks.
}.
Two colors (red and green, say)
can pair via gluon interactions in the
(\thrbarA,\thrS,1,$+$)($C\ga_3$) channel, and the
 blue $s$ quarks can then pair in the rotationally invariant
(\sixS,\thrS,0,$+$)($C\ga_0\ga_5$). Note that these pairing patterns,
being flavor symmetric, only exist for the single-gluon interaction,
not the instanton interaction, so their strength
is model-dependent.
This leaves the blue $u$ quarks and $d$ quarks. We have not
found a single-color and single-flavor channel that
can pair massless quarks, so they will have to find
an attractive channel outside the set that we have explored here.
An alternative possibility is that for each flavor
there is color-spin locked (CSL) condensation
of all three colors.

\item Two-flavor $u$-$d$ and additional single flavor pairing (2SC+1SC).\\
It is just possible that there is a range of $m_s$ in which 2SC $u$-$d$
pairing can occur. $m_s$ must be high enough so that CFL is
disfavored relative to 2SC, but low enough so that the $u$ and $d$
Fermi momenta are close enough to allow them to pair with each other.
The energy accounting is complicated by contributions from the chiral
condensate, which competes more or less strongly with different quark
pairing patterns. Current indications are that
2SC either does not occur, or survives only in a very narrow window
\cite{AR-02,Steiner:2002gx}.
The pairings involved in 2SC+1SC are depicted in Fig.~\ref{fig:2sc_levels}. 
The red and green $u$ and $d$ quarks undergo 2SC pairing.
The blue quarks and the
strange quarks follow the same pairing patterns as in the 
single-flavor pairing (1SC) case.
Note that enforcing color neutrality will create
a small $\O(\De^2/\mu)$ splitting between
the Fermi momentum of the blue $s$ and 
that of the the red and green $s$. This may
prevent color-spin-locking for the $s$ quarks.

\item Three-flavor crystalline pairing (LOFF). \\
An important competitor to the
channels discussed here is the LOFF phase, in which flavors that
cannot undergo conventional BCS pairing because their Fermi momenta
are too far apart can nevertheless pair over part of their Fermi
surfaces by forming pairs with net momentum. This breaks translational
invariance, and leads to a crystal structure \cite{LOFF,OurLOFF}.
One can imagine a type of CFL-LOFF pairing, in which the red and green
$u$ and $d$ quarks, the red and blue $u$ and $s$ quarks,
and the blue and green $d$ and $s$ quarks, all form LOFF pairs.
A recent Ginzburg-Landau study \cite{Bowers:2002xr} indicates
that the condensation energy of the crystalline phase may be large
enough to beat any single-flavor channel,
although this result followed from extrapolating the Ginzburg-Landau analysis
beyond its realm of validity. 

\een

As well as evaluating the attractiveness of a large set of channels,
we have calculated the angular dependence over the Fermi surface of the
energies of the quasiquark excitations associated with each of the
$j=1$ quark pair condensates. 
We see that the $j=1,m_j=\pm 1$ condensates always have gapless
quasiquarks at the poles of the Fermi sphere.
The $j=1,m_j=0$ condensates have gapless regions
for massless quarks, at the poles ($\psi C\ga_3\psi$) or around the
equator ($\psi C\si_{03}\psi$), but if the quarks are massive then the
the quasiquarks have a minimum gap of order $m\De/\mu$
(Figs.~\ref{fig:DRmassless},\ref{fig:DRmassive}). The
gapless points or lines on the Fermi surface are likely to lead
to interesting phenomenology (see below).

Finally, we have surveyed the isotropic mesonic condensates,
and find that the conventional chiral condensate is very heavily favored.
The color-flavor-locked mesonic channel studied in Ref.~\cite{zeroden}
is strongly disfavored, as was noted in Ref.~\cite{Berges:2000nb}.
It has been argued \cite{Wetterich:2000ky} that taking into account the
effect of the color-flavor-locked channel on the
integral over instanton sizes makes the condensation energy
come out attractive after all.

\section{The road ahead}
This work can be extended in many directions, and points
to interesting questions in the theory of dense matter and compact star
phenomenology.\\
{\bf Dense matter:}\\
(1) In order to allow comparisons between the phases we studied and other
competing possibilities such as the LOFF phase, it would be useful to
calculate the free energy of the various 1SC pairing patterns.
(2) It would be valuable to extend our survey to include
an even wider set of possible quark pair condensation patterns,
such as color-spin locking. We would also like to find
a channel in which light single-color single-flavor
pairing can occur. It might be necessary to explore
condensates where orbital angular momentum is introduced explicitly 
via derivatives: $\psi C \p_\mu \psi$ etc. These momentum-dependent
condensates will not arise from the mean-field treatment
of the simplified point-like interactions 
that we used. One would have to give the four-fermion vertex a
non-trivial form factor, or go beyond the mean-field approximation.
(3) A related area that is already being investigated is
single color/flavor channels
in the ultra-high-density region with the true QCD
interaction vertex, including the hard-dense-loop-resummed gluon
propagator \cite{TS1flav,rockefeller,Schmitt:2002sc}. 
In this regime one will also find momentum-dependent gaps.
(4) We studied the attractive channels independently of each other, and
independent of chiral condensation. It is clearly
necessary to solve
their coupled gap equations, and see how they compete/coexist.
(5) We only worked at zero temperature. It is straightforward to
generalize the gap equations to arbitrary temperature, and recent work
indicates that the usual BCS relationship between the critical temperature
and the gap parameter may be modified for these exotic condensates
\cite{BHO,Schmitt:2002sc}.

\noindent {\bf Compact star phenomenology:}\\
Although many of the channels we studied have very small gaps
and therefore very small critical temperatures, they 
will be phenomenologically relevant if they are the best pairing option
available for some of the quarks. Since the temperature of a compact star
falls to tens of eV when its age reaches about a million years, 
pairing can suddenly occur in such channels late in the star's life,
and the corresponding quasiquark excitations will suddenly become
too heavy to participate in transport processes.
We must remember, however, that some of the exotic channels are
gapless at special 3-momenta. In those cases
a small proportion of their quasiparticles may continue to
play a role in transport properties, even
when the temperature is much less than the gap parameter. Some specific
topics for further investigation are:
(1) Develop the transport theory
of $j\neq 0$ condensates: they have gapless
modes at points or along lines on their Fermi surface.
What effect does this have on
neutrino emission/absorption via URCA processes or otherwise,
their specific heat, viscosities, conductivities, etc?
A natural first step would be to write down an effective theory,
which would contain the lowest quasiquark modes, Goldstone bosons
arising from the breaking of rotational symmetry (which could be called 
"spin waves" by analogy with $^3$He), 
and density waves.
(2) The $j=1$ condensates can carry angular momentum simply by
aligning themselves in large domains, without involving
any superfluid vortices, but it seems they will typically occur
in conjunction with other phases that are superfluid.
It would be interesting to see how the angular momentum is
carried in this situation.
(3) The  $j=1$ condensates are ferromagnets.
It would be interesting to see if they could
generate large magnetic fields such as those suggested
for magnetars ($10^{14-15}$~Gauss).
(4) The various $j=1$ condensates show completely different 
variation of the energy gap over the Fermi surface
(Section~\ref{sec:dispersion}).
It would be useful to know how they behave when exposed
to the nonisotropic external influences that are common in
compact stars, such as magnetic fields
or neutrino fluxes, and also in their coupling to external
sources of torque, eg via electron-quasiquark scattering.

Clearly there remain many interesting questions, both formal and 
phenomenological, about the single
color and/or single flavor color superconducting phases
of dense quark matter.

\vspace{3ex}
{\samepage 
\begin{center} Acknowledgements \end{center}
\nopagebreak
The work of MGA and GAC is supported by the UK PPARC.
The work of JAB is supported by the U.S. Department
of Energy (D.O.E.) under cooperative research agreement \#DF-FC02-94ER40818
and by a DOD National Defense Science and Engineering Graduate Fellowship. 
The work of JMC is supported by a University Scholarship from
Glasgow University, and additional funding from the Glasgow University 
physics department.
MGA and JAB acknowledge the Kavli Institute for
Theoretical Physics for providing a stimulating venue for part of
this work.
We thank Krishna Rajagopal for very valuable discussions, and
Thomas Sch\"afer for comments on an earlier draft.
}

\appendix
\renewcommand{\theequation}{\Alph{section}.\arabic{equation}}

\section{Calculational details}
\label{app:gap}

To allow the possibility of quark pairing, we use
8-component Nambu-Gorkov spinors,
\beq
\Psi=\left(\begin{tabular}{c}$\psi(p)$ \\ $\bar{\psi}^T(-p)$
 \end{tabular}\right)
\eeq
with
\beq
\bar{\Psi}=\left(\bar{\psi}(p),\psi^T(-p)\right)~.
\eeq
In Minkowski space
the inverse quark propagator for massive fermions takes the form,
\beq
S^{-1}(p)=\Bigg(\begin{tabular}{cc}$\pslash-m+\mu\gamma_0$ & $\bar{\Delta}$\\
				  $\Delta$ & $(\pslash+m-\mu\gamma_0)^T$
	       \end{tabular}\Bigg)
\label{invprop}
\eeq
where
\beq
\bar{\Delta}=\gamma_0\Delta^\dagger\gamma_0~.
\eeq
The gap matrix $\Delta$ is a matrix in color, flavor and Dirac space,
multiplied by a gap parameter also denoted as $\De$,
\beq
\Delta^{\alpha\beta ab}_{ij} = \Delta(\mu) {\cal C}^{\rm \alpha\beta}
  {\cal F}_{ij}\Gamma^{ab}~.
\eeq
The relation between the proper self energy and the full propagator is,
\beq
S^{-1}=S_0^{-1}+\Sigma=\Bigg(\begin{tabular}{cc}$\pslash-m+\mu\gamma_0$ & 0\\
				  0 & $(\pslash+m-\mu\gamma_0)^T$
	       \end{tabular}\Bigg)+
		\Bigg(\begin{tabular}{cc} 0 & $\bar{\Delta}$\\
				  $\Delta$ & 0
	       \end{tabular}\Bigg)
\eeq
where $S_0^{-1}$ is the inverse propagator in the absence of interactions.
The gap is determined by solving a self-consistent Schwinger-Dyson equation for
$\Sigma$. For a 4-fermion interaction modelling single gluon exchange, this
takes the form
\beq
\Sigma=-6iG\int\frac{d^4p}{(2\pi)^4}V^A_\mu S(p)V^{A\mu}
\label{gapeq}
\eeq
where $V^A_\mu$ 
is the interaction vertex in the Nambu-Gorkov basis. 
We study three interactions, the quark-gluon vertex
\beq
V^A_\mu=\Bigg(\begin{tabular}{cc}$\gamma_\mu\lambda^A/2$ & 0\\
				  0 & $-(\gamma_\mu\lambda^A/2)^T$
	       \end{tabular}\Bigg)~,
\eeq
the quark-magnetic gluon vertex
\beq
V^A_i=\Bigg(\begin{tabular}{cc}$\gamma_i\lambda^A/2$ & 0\\
				  0 & $-(\gamma_i\lambda^A/2)^T$
	       \end{tabular}\Bigg)~,
\eeq
and the quark-instanton vertex, for which
\beq
\Sigma_{ik}^{\al\ga}=-6iG\int\frac{d^4p}{(2\pi)^4} \Bigl(
 V_{L\mu}^A S_{jl}^{\be\de}(p)V_L^{A\mu} + V_{R\mu}^A S_{jl}^{\be\de}(p)V_R^{A\mu} \Bigr)\Xi_{ik\be\de}^{jl\al\ga}
\eeq
where
\beq
\Xi_{ik\be\de}^{jl\al\ga}=-\ep_{ik}\ep^{jl}\frac{2}{3} (3 \de^\al_\be \de^\ga_\de - \de^\al_\de \de^\ga_\be)
\eeq
and
\beq
V_L^A=\Bigg(\ba{cc}({\mathbf 1}+\gamma_5) & 0\\
		 0 & ({\mathbf 1}+\gamma_5)^T
	       \ea\Bigg)~,
\quad\hbox{and}\quad
V_R^A=\Bigg(\ba{cc}({\mathbf 1}-\gamma_5) & 0\\
		 0 & ({\mathbf 1}-\gamma_5)^T
	       \ea\Bigg)~.
\eeq
In the case of the $\psi C\ga_5\psi$ condensate for the full gluon interaction
we obtain the gap equation, which after rotation to Euclidean space becomes
\beq
1=16G\int\frac{d{p_0}d^3p}{(2\pi)^4}
  \frac{4(\Delta^2+\mu^2+{p_0}^2+p^2)}{{W}}
\eeq
where
\beq
{W}=\Delta^4+\mu^4+({p_0}^2+p^2)^2+2\Delta^2(\mu^2+{p_0}^2+p^2)-2\mu^2(-{p_0}^2+p^2)~.
\eeq
The $p_0$ integral can be explicitly evaluated,
\beq
1=\frac{2G}{\pi^2}\int^\Lambda_0 dp\left[
   \frac{p^2}{\sqrt{\Delta^2+(p+\mu)^2}}
  +\frac{p^2}{\sqrt{\Delta^2+(p-\mu)^2}}\right]~.
\eeq
The momentum integral can be performed
analytically, giving
\beq
\Delta=2\sqrt{\Lambda^2-\mu^2}\exp\left(
  \frac{\Lambda^2-3\mu^2}{2\mu^2}\right)\exp\left(-\frac{\pi^2}{4\mu^2G}\right)
\label{analyticgap}
\eeq
for $\De\ll\mu$.

\section{Gap Equation Summary}

Here are the gap equations for the attractive channels. In the
following, positive square roots are implied and we 
define $p_r^2\equiv(p_x)^2+(p_y)^2$.

$$ \int d|p|\equiv \int^\La_{0} d|p|,\quad
  \int dp_r dp_z \equiv \int^\La_0 dp_r \int^{\sqrt{\La^2-p_r^2}}_{-\sqrt{\La^2-p_r^2}} dp_z
$$

\subsection{$C\ga_5$ and $C$ gap equations}

\begin{equation}
1=N\frac{G}{\pi^2}\int d|p|\; \left[
   \frac{|p|^2}{\sqrt{\De^2+(|p|-\mu)^2}}
  +\frac{|p|^2}{\sqrt{\De^2+(|p|+\mu)^2}}
  \right]
\end{equation}
where N is a constant that differs for each interaction.
$$
\begin{array}{lc}
\mbox{Instanton} &\hspace{2cm} N=4 \\
\mbox{Magnetic + Electric Gluon} &\hspace{2cm} N=2 \\
\mbox{Magnetic Gluon} &\hspace{2cm} N=\frac{3}{2} 
\end{array}
$$

The $C$ channel produces an identical gap equation for both the full
gluon and magnetic gluon interactions. The instanton interaction is
not attractive in this channel.

\subsection{$C\sigma_{03}$ and $C\sigma_{03}\ga_5$ gap equations}

\begin{equation}
1=N\frac{G}{\pi^2}\int dp_r \;dp_z \: \left[
  \frac{p_r(\mathcal{E}+p_r^2)}{\mathcal{E} E_+}
 +\frac{p_r(\mathcal{E}-p_r^2)}{\mathcal{E} E_-} 
 \right]
\end{equation}
with
\begin{eqnarray*}
\mathcal{E}^2&=& \De^2 p_r^2 + \mu^2|p|^2 \\
E_{\pm}^2&=&\De^2 + \mu^2 + |p|^{2} \pm 2\mathcal{E} 
\end{eqnarray*}
where N is a constant that differs for each interaction.
$$
\begin{array}{lc}
\mbox{Instanton} &\hspace{2cm} N=1 \\
\mbox{Magnetic Gluon} &\hspace{2cm} N=\frac{1}{8} 
\end{array}
$$

The $C\sigma_{03}\ga_5$ channel produces an identical gap equation for
magnetic gluon interaction. The instanton and the magnetic gluon
interactions are not attractive in this channel.

\subsection{$C(\sigma_{01} \pm i \sigma_{02})$ gap equation}

\begin{equation}
1=N\frac{-iG}{\pi^3}\int dp_r\;dp_z \int^\infty_{-\infty} dp^0\: 
  \frac{p_r (\mu^2-(p^0)^2-p_z^2-2p^0p_z) }{W}
\end{equation}
where
$$W=\mu^4+(-(p^0)^2+|p|^2)^2+2 \De^2(\mu^{2}-(p^0)^2-p_z^2-2p^0p_z)-2 \mu^2((p^0)^2+|p|^2)$$
and N is a constant that differs for each interaction.
$$
\begin{array}{lc}
\mbox{Instanton} &\hspace{2cm} N=2 \\
\mbox{Magnetic Gluon} &\hspace{2cm} N=\frac{1}{4} 
\end{array}
$$

\subsection{$C\ga_{3}$ gap equation}

\begin{equation}
1=N\frac{G}{\pi^2}\int dp_r \;dp_z \: \left[
  \frac{p_r(\mathcal{E}+p_z^2)}{\mathcal{E} E_+}
 +\frac{p_r(\mathcal{E}-p_z^2)}{\mathcal{E} E_-} 
 \right]
\end{equation}
with
\begin{eqnarray*}
\mathcal{E}^2&=& \De^2 p_z^2 + \mu^2(|p|^2+m^2) \\
E_{\pm}^2&=&\De^2 + \mu^2 +m^2 + |p|^{2} \pm 2\mathcal{E} 
\end{eqnarray*}
where N is a constant that differs for each interaction.
$$
\begin{array}{lc}
\mbox{Magnetic + Electric Gluon} &\hspace{2cm} N=\frac{1}{2} \\
\mbox{Magnetic Gluon} &\hspace{2cm} N=\frac{1}{4} 
\end{array}
$$

\subsection{$C\ga_{3}\ga_5$ gap equation}

This channel is not attractive for instantons and the magnetic gluon
interaction is the same as for the massless $C\ga_{3}$ channel, i.e.

\begin{equation}
1=N\frac{G}{\pi^2}\int dp_r \;dp_z \: \left[
  \frac{p_r(\mathcal{E}+p_z^2)}{\mathcal{E} E_+}
 +\frac{p_r(\mathcal{E}-p_z^2)}{\mathcal{E} E_-} 
 \right]
\end{equation}
with
\begin{eqnarray*}
\mathcal{E}^2&=& \De^2 p_z^2 + \mu^2|p|^2 \\
E_{\pm}^2&=&\De^2 + \mu^2 + |p|^{2} \pm 2\mathcal{E} 
\end{eqnarray*}

\subsection{$C (\ga_1 \pm i \ga_2)$ gap equation}

\begin{equation}
1=N\frac{-iG}{\pi^3}\int dp_r\;dp_z \int^\infty_{-\infty} dp^0\: 
  \frac{p_r (2\De^2 E_{1-} E_{1+}+(m^2+\mu^2-(p^0)^2+p_z^2) E_{2-} E_{2+})}{
    E_{1-} E_{1+}(4\De^4 + 4\De^2(m^2+\mu^2-(p^0)^2+p_z^2) + E_{1-} E_{1+})} 
\end{equation}
with
\begin{eqnarray*}
E_{1\pm}&=&m^2-(\mu\pm p^0)^2+p_z^2 \\
E_{2\pm}&=&m^2-(\mu\pm p^0)^2+|p|^2 
\end{eqnarray*}
where N is a constant that differs for each interaction.
$$
\begin{array}{lc}
\mbox{Magnetic + Electric Gluon} &\hspace{2cm} N=1 \\
\mbox{Magnetic Gluon} &\hspace{2cm} N=\frac{1}{2} 
\end{array}
$$

\subsection{$C\ga_0\ga_5$ gap equation}

\begin{equation}
1=N\frac{G}{\pi^2}\int d|p| \: \left[
  \frac{|p|^2(\mathcal{E}+|p|^2)}{\mathcal{E} E_+}
 +\frac{|p|^2(\mathcal{E}-|p|^2)}{\mathcal{E} E_-} 
 \right]
\end{equation}
with
\begin{eqnarray*}
\mathcal{E}^2&=& \De^2 |p|^2 + \mu^2(|p|^2+m^2) \\
E_{\pm}^2&=&\De^2 + \mu^2 +m^2 + |p|^{2} \pm 2\mathcal{E} 
\end{eqnarray*}
where N is a constant that differs for each interaction.
$$
\begin{array}{lc}
\mbox{Magnetic + Electric Gluon} &\hspace{2cm} N=\frac{1}{2} \\
\mbox{Magnetic Gluon} &\hspace{2cm} N=\frac{3}{4} 
\end{array}
$$
For $m=0$ this reduces to 
\begin{equation}
1=N\frac{2G}{3\pi^2}\frac{\La^3}{\sqrt{\De^2+\mu^2}}
\end{equation}

\section{Orbital/spin content of the condensates}
\label{app:angmom}

In the non-relativistic limit it is meaningful to ask about the
separate contributions of the orbital and spin angular momenta
to the total angular momentum of the diquark condensates.
We can find out by expanding the
field operators out of which the condensates are built in terms
of creation and annihilation operators,
\beq
\psi^\alpha_i=\sum_{k,s}\left(\frac{m}{VE_k}\right)^\frac{1}{2}
  \left[u^s(k)a^s_{ki\alpha}e^{-ikx}
       +v^s(k)b^{s\dagger}_{ki\alpha}e^{ikx}\right]
\eeq
Inserting the explicit momentum-dependent spinors in any basis allows
the creation/annihilation operator expansions of the condesates to be
calculated.

In the Dirac basis,
\beq
\ba{rcl@{\qquad}rcl}
u_1^D(\vk) &=& A\left(
 \ba{c} 1 \\ 0 \\ Bk_3 \\ B(k_1+ik_2)\ea\right)
&
u_2^D(\vk) &=& A\left(
 \ba{c} 0 \\ 1 \\ B(k_1-ik_2) \\ -Bk_3\ea\right)\\[6ex]
v_1^D(\vk) &=& A\left(
 \ba{c} B(k_1-ik_2) \\ -Bk_3 \\ 0 \\ 1\ea\right)
&
v_2^D(\vk) &=& A\left(
 \ba{c} Bk_3 \\ B(k_1+ik_2) \\ 1 \\ 0\ea\right)
\ea
\eeq
\beq
A=\left(\frac{E+m}{2m}\right)^\half, \qquad B=\frac{1}{E+m}
\eeq

Eq.~\eqn{eqn:C_op_exp} shows the result of performing such a 
calculation for the $\psi C \psi$ condensate,
\bea
\nonumber
\label{eqn:C_op_exp}
\conds{C}&=&\frac{1}{E}\left[(a^2_{pi\alpha}a^2_{-pj\beta}
  +b^{\dagger 1}_{pi\alpha}b^{\dagger 1}_{-pj\beta})(p_1-ip_2)
  \right.\\\nonumber\\\nonumber
&\phantom{=}&-(a^1_{pi\alpha}a^1_{-pj\beta}
  +b^{\dagger 2}_{pi\alpha}b^{\dagger 2}_{-pj\beta})(p_1+ip_2)\\\nonumber\\
&\phantom{=}&\left.+2(a^1_{pi\alpha}a^2_{-pj\beta}
  +b^{\dagger 1}_{pi\alpha}b^{\dagger 2}_{-pj\beta})p_3\right]\cfms
\eea
where $\cfms$ is the
color-flavor matrix which is symmetric under the
interchange $i\rightleftharpoons j$ (flavor) and 
$\alpha\rightleftharpoons\beta$
(color). A sum over momentum $p$ should be performed on the right hand 
side.

Once the operator expansions have been obtained it is a relatively simple
procedure to obtain the angular momentum content by rearranging the terms
and inserting the relevant spherical harmonics. It is important to include 
contributions from momenta $k$ and $-k$ together, since they involve the
same creation/annihilation operators.
For example, for the condensate in \eqn{eqn:C_op_exp},
\bea
p=k&:&\frac{1}{E}\left[a^2_{ki\alpha}a^2_{-kj\beta}(k_1-ik_2)
-a^1_{ki\alpha}a^1_{-kj\beta}(k_1+ik_2)
+2a^1_{ki\alpha}a^2_{-kj\beta}k_3\right]\\
p=-k&:&\frac{1}{E}\left[-a^2_{-ki\alpha}a^2_{kj\beta}(k_1-ik_2)
+a^1_{-ki\alpha}a^1_{kj\beta}(k_1+ik_2)
-2a^1_{-ki\alpha}a^2_{kj\beta}k_3\right]\\\nonumber
&\to&\frac{1}{E}\left[a^2_{ki\alpha}a^2_{-kj\beta}(k_1-ik_2)
-a^1_{ki\alpha}a^1_{-kj\beta}(k_1+ik_2)
+2a^2_{ki\alpha}a^1_{-kj\beta}k_3\right]
\eea
where we have relabelled $k\to -k, i\leftrightarrow j, 
\al \leftrightarrow \be$ in the last line.
The final result is a sum over $k,\al,\be,i,j$ of
\beq
\frac{2}{E}\left[a^2_{ki\alpha}a^2_{-kj\beta}(k_1-ik_2)
-a^1_{ki\alpha}a^1_{-kj\beta}(k_1+ik_2)
+(a^1_{ki\alpha}a^2_{-kj\beta}+a^2_{ki\alpha}a^1_{-kj\beta})k_3\right]
\eeq
Upon inserting the relevant spherical harmonics and using standard
arrow notation for the spins we obtain
\beq
\conds{C}\rightarrow-2\sqrt{8\pi} \frac{p}{E}\left[
  \frac{1}{\sqrt{3}}\uu Y_1^{-1}
 +\frac{1}{\sqrt{3}}\dd Y_1^1
 -\frac{1}{\sqrt{3}}\frac{1}{\sqrt{2}}[\ud+\du]Y_1^0
 \right]
\eeq
which has precisely the correct Clebsch-Gordan structure to be interpreted
as a state with orbital angular momentum $l=1$, which gives an 
antisymmetric spatial wave function, and spin $s=1$, which gives
a symmetric spin wavefunction, combined to give $j=0$. We write this
as $ |l=1_A,s=1_S\rangle$.
Applying this to all the condensates we studied, we can make a table of the
particle-particle (as opposed to particle-hole) content of each of them,
\beq
\label{BCSenhancement}
\ba{c@{\qquad}rl}
\conds{C\gamma_5} & 4\sqrt{2\pi} & | l=0_S,s=0_A\rangle\\[1ex]
\conds{C} & \dsp -2\sqrt{8\pi}\frac{p}{E} & | l=1_A,s=1_S\rangle\\[2ex]
\conds{C\gamma_0\gamma_5} 
  & \dsp 4\sqrt{2\pi}\frac{m}{E} & | l=0_S,s=0_A\rangle\\[2ex]
\conds{C\gamma_3\gamma_5}
  & \dsp 8\sqrt{\frac{\pi}{3}}\frac{p}{E} & | l=1_A,s=1_S\rangle\\[2ex]
\conda{C\sigma_{03}\gamma_5}
& \dsp -4i\sqrt{\frac{2\pi}{3}}\frac{p}{E} & | l=1_A,s=0_A\rangle\\[2ex]
\conda{C\sigma_{03}}
  & \multicolumn{2}{r}{ \left\{
    \ba{r} \dsp -\frac{8}{3}\sqrt{\pi}i\frac{p^2}{E(E+m)} 
    | l=2_S,s=1_S\rangle\\[3ex]
   \dsp + 2i\sqrt{2\pi}\left[\frac{(E+m)^2-\frac{1}{3}p^2}{E(E+m)}\right]
    | l=0_S,s=1_S\rangle
    \ea \right. }  \\[8ex]
\conda{C\gamma_0}& 0 \\[4ex]
\conda{C\gamma_3}
  &  \multicolumn{2}{r}{ \left\{
   \ba{r} \dsp \frac{8}{3}\sqrt{\pi}\frac{p^2}{E(E+m)} 
    | l=2_S,s=1_S\rangle\\[3ex]
    \dsp +2\sqrt{2\pi}\left[\frac{(E+m)^2-\frac{1}{3}p^2}{E(E+m)}\right] 
    |  l=0_S,s=1_S\rangle
    \ea \right. }  \\
\ea
\eeq
These results are summarized in the ``BCS-enhanced'' 
column of Table~\ref{tab:channels}.
We can see explicitly that the $\psi C \ga_0\ga_5 \psi$ condensate
has no particle-particle component in the massless limit, which is why
it cannot occur at high density for the up and down quarks.
This reflects basic physics: the condensate has spin zero, so the two spins
must be oppositely aligned. But it is an ``LR'' condensate 
(see Table~\ref{tab:channels}), so in the massless limit the two quarks,
having opposite momentum and opposite helicity, have parallel spins.

\end{document}